\documentclass[submission,copyright]{eptcs}
\usepackage{etex} % keep just after \documentclass

\providecommand{\event}{QPL 2015} % Name of the event you are submitting to
\newcommand{\urlalt}[2]{\href{#1}{\urlstyle{rm}\nolinkurl{#2}}}

\newif\ifignore % when set to true, additional text appears containing
\ignorefalse

\newcommand{\auxproof}[1]{%
\ifignore\mbox{}\newline
\textbf{PROOF:} \dotfill\newline
{\it #1}\mbox{}\newline
\textbf{ENDPROOF}\dotfill\newline
\fi}

\usepackage[english]{babel}
\usepackage{amsmath}
\usepackage{enumitem}
\usepackage{amssymb}
\usepackage{xspace}
\usepackage{stmaryrd}
\usepackage{picins}
\usepackage{url,hyperref}
\usepackage{fancybox}%\usepackage{rotating}
\usepackage{xy}
\usepackage{mathrsfs}
\xyoption{all} % Really need v2?
\xyoption{2cell}
\UseTwocells

\usepackage{tikz}

\newdimen\proofrulebreadth \proofrulebreadth=.05em
\newdimen\proofdotseparation \proofdotseparation=1.25ex
\newdimen\proofrulebaseline \proofrulebaseline=2ex
\newcount\proofdotnumber \proofdotnumber=3
\let\then\relax
\def\hfi{\hskip0pt plus.0001fil}
\mathchardef\squigto="3A3B
\newif\ifinsideprooftree\insideprooftreefalse
\newif\ifonleftofproofrule\onleftofproofrulefalse
\newif\ifproofdots\proofdotsfalse
\newif\ifdoubleproof\doubleprooffalse
\let\wereinproofbit\relax
\newdimen\shortenproofleft
\newdimen\shortenproofright
\newdimen\proofbelowshift
\newbox\proofabove
\newbox\proofbelow
\newbox\proofrulename
\def\shiftproofbelow{\let\next\relax\afterassignment\setshiftproofbelow\dimen0 }
\def\shiftproofbelowneg{\def\next{\multiply\dimen0 by-1 }%
\afterassignment\setshiftproofbelow\dimen0 }
\def\setshiftproofbelow{\next\proofbelowshift=\dimen0 }
\def\setproofrulebreadth{\proofrulebreadth}

\def\prooftree{% NESTED ZERO (\ifonleftofproofrule)
\ifnum  \lastpenalty=1
\then   \unpenalty
\else   \onleftofproofrulefalse
\fi
\ifonleftofproofrule
\else   \ifinsideprooftree
        \then   \hskip.5em plus1fil
        \fi
\fi
\bgroup% NESTED ONE (\proofbelow, \proofrulename, \proofabove,
\setbox\proofbelow=\hbox{}\setbox\proofrulename=\hbox{}%
\let\justifies\proofover\let\leadsto\proofoverdots\let\Justifies\proofoverdbl
\let\using\proofusing\let\[\prooftree
\ifinsideprooftree\let\]\endprooftree\fi
\proofdotsfalse\doubleprooffalse
\let\thickness\setproofrulebreadth
\let\shiftright\shiftproofbelow \let\shift\shiftproofbelow
\let\shiftleft\shiftproofbelowneg
\let\ifwasinsideprooftree\ifinsideprooftree
\insideprooftreetrue
\setbox\proofabove=\hbox\bgroup$\displaystyle % NESTED TWO
\let\wereinproofbit\prooftree
\shortenproofleft=0pt \shortenproofright=0pt \proofbelowshift=0pt
\onleftofproofruletrue\penalty1
}

\def\eproofbit{% NESTED TWO
\ifx    \wereinproofbit\prooftree
\then   \ifcase \lastpenalty
        \then   \shortenproofright=0pt  % 0: some other object, no indentation
        \or     \unpenalty\hfil         % 1: empty hypotheses, just glue
        \or     \unpenalty\unskip       % 2: just had a tree, remove glue
        \else   \shortenproofright=0pt  % eh?
        \fi
\fi
\global\dimen0=\shortenproofleft
\global\dimen1=\shortenproofright
\global\dimen2=\proofrulebreadth
\global\dimen3=\proofbelowshift
\global\dimen4=\proofdotseparation
\global\count255=\proofdotnumber
$\egroup  % NESTED ONE
\shortenproofleft=\dimen0
\shortenproofright=\dimen1
\proofrulebreadth=\dimen2
\proofbelowshift=\dimen3
\proofdotseparation=\dimen4
\proofdotnumber=\count255
}

\def\proofover{% NESTED TWO
\eproofbit % NESTED ONE
\setbox\proofbelow=\hbox\bgroup % NESTED TWO
\let\wereinproofbit\proofover
$\displaystyle
}%
\def\proofoverdbl{% NESTED TWO
\eproofbit % NESTED ONE
\doubleprooftrue
\setbox\proofbelow=\hbox\bgroup % NESTED TWO
\let\wereinproofbit\proofoverdbl
$\displaystyle
}%
\def\proofoverdots{% NESTED TWO
\eproofbit % NESTED ONE
\proofdotstrue
\setbox\proofbelow=\hbox\bgroup % NESTED TWO
\let\wereinproofbit\proofoverdots
$\displaystyle
}%
\def\proofusing{% NESTED TWO
\eproofbit % NESTED ONE
\setbox\proofrulename=\hbox\bgroup % NESTED TWO
\let\wereinproofbit\proofusing
\kern0.3em$
}

\def\endprooftree{% NESTED TWO
\eproofbit % NESTED ONE
  \dimen5 =0pt% spread of hypotheses
\dimen0=\wd\proofabove \advance\dimen0-\shortenproofleft
\advance\dimen0-\shortenproofright
\dimen1=.5\dimen0 \advance\dimen1-.5\wd\proofbelow
\dimen4=\dimen1
\advance\dimen1\proofbelowshift \advance\dimen4-\proofbelowshift
\ifdim  \dimen1<0pt
\then   \advance\shortenproofleft\dimen1
        \advance\dimen0-\dimen1
        \dimen1=0pt
        \ifdim  \shortenproofleft<0pt
        \then   \setbox\proofabove=\hbox{%
                        \kern-\shortenproofleft\unhbox\proofabove}%
                \shortenproofleft=0pt
        \fi
\fi
\ifdim  \dimen4<0pt
\then   \advance\shortenproofright\dimen4
        \advance\dimen0-\dimen4
        \dimen4=0pt
\fi
\ifdim  \shortenproofright<\wd\proofrulename
\then   \shortenproofright=\wd\proofrulename
\fi
\dimen2=\shortenproofleft \advance\dimen2 by\dimen1
\dimen3=\shortenproofright\advance\dimen3 by\dimen4
\ifproofdots
\then
        \dimen6=\shortenproofleft \advance\dimen6 .5\dimen0
        \setbox1=\vbox to\proofdotseparation{\vss\hbox{$\cdot$}\vss}%
        \setbox0=\hbox{%
                \advance\dimen6-.5\wd1
                \kern\dimen6
                $\vcenter to\proofdotnumber\proofdotseparation
                        {\leaders\box1\vfill}$%
                \unhbox\proofrulename}%
\else   \dimen6=\fontdimen22\the\textfont2 % height of maths axis
        \dimen7=\dimen6
        \advance\dimen6by.5\proofrulebreadth
        \advance\dimen7by-.5\proofrulebreadth
        \setbox0=\hbox{%
                \kern\shortenproofleft
                \ifdoubleproof
                \then   \hbox to\dimen0{%
                        $\mathsurround0pt\mathord=\mkern-6mu%
                        \cleaders\hbox{$\mkern-2mu=\mkern-2mu$}\hfill
                        \mkern-6mu\mathord=$}%
                \else   \vrule height\dimen6 depth-\dimen7 width\dimen0
                \fi
                \unhbox\proofrulename}%
        \ht0=\dimen6 \dp0=-\dimen7
\fi
\let\doll\relax
\ifwasinsideprooftree
\then   \let\VBOX\vbox
\else   \ifmmode\else$\let\doll=$\fi
        \let\VBOX\vcenter
\fi
\VBOX   {\baselineskip\proofrulebaseline \lineskip.2ex
        \expandafter\lineskiplimit\ifproofdots0ex\else-0.6ex\fi
        \hbox   spread\dimen5   {\hfi\unhbox\proofabove\hfi}%
        \hbox{\box0}%
        \hbox   {\kern\dimen2 \box\proofbelow}}\doll%
\global\dimen2=\dimen2
\global\dimen3=\dimen3
\egroup % NESTED ZERO
\ifonleftofproofrule
\then   \shortenproofleft=\dimen2
\fi
\shortenproofright=\dimen3
\onleftofproofrulefalse
\ifinsideprooftree
\then   \hskip.5em plus 1fil \penalty2
\fi
}

\usepackage{mathtools}
\usepackage{nicefrac}

\newtheorem{theorem}{\textbf{Theorem}}
\newtheorem{lemma}[theorem]{\textbf{Lemma}}
\newtheorem{proposition}[theorem]{\textbf{Proposition}}
\newtheorem{corollary}[theorem]{\textbf{Corollary}}
\newtheorem{definition}[theorem]{\textbf{Definition}}

\newtheorem{notation}[theorem]{Notation}

\newtheorem{remark}[theorem]{Remark}
\newenvironment{myproof}[1][Proof]%
   { \begin{trivlist}%
     \item[\hskip \labelsep {\bfseries #1}]%
   }%
   { \end{trivlist}%
   }

\newcommand{\keyword}[1]{\textbf{#1}}
\newcommand{\after}{\mathbin{\circ}}
\newcommand{\set}[2]{\{#1\;|\;#2\}}
\newcommand{\setin}[3]{\{#1\in#2\;|\;#3\}}

\newcommand{\allin}[3]{\forall{#1\in#2}.\,#3}

\newcommand{\downset}{\mathop{\downarrow}}

\newcommand{\C}{\mathbb{C}}
\newcommand{\Z}{\mathbb{Z}}
\newcommand{\intd}{{\kern.2em}\mathrm{d}{\kern.03em}}
\newcommand{\ket}[1]{\ensuremath{|{\kern.1em}#1{\kern.1em}\rangle}}

\newcommand{\ie}{\textit{i.e.}\xspace}

\newcommand{\Dst}{\mathcal{D}}

\newcommand{\Giry}{\mathcal{G}}
\newcommand{\Gsub}{\Giry_{\le1}}

\newcommand{\Pow}{\mathcal{P}}
\newcommand{\Clopen}{\mathrm{Clopen}}
\newcommand{\MeasS}{\mathrm{Meas}}

\newcommand{\nePow}{\Pow_{*}}
\newcommand{\idmap}[1][]{\ensuremath{\mathrm{id}_{#1}}}

\newcommand{\op}[1]{#1^{\textrm{op}}}

\newcommand{\supp}{\textsl{supp}}

\newcommand{\ev}{\textsl{ev}}

\newcommand{\indic}[1]{\mathbf{1}_{#1}}

\newcommand{\floor}[1]{\ensuremath{\lfloor#1\rfloor}}
\newcommand{\ceil}[1]{\ensuremath{{\lceil#1\rceil}}}
\newcommand{\cmpr}[2]{\ensuremath{\{#1|{\kern.2ex}#2\}}}
\newcommand{\instr}{{\textrm{instr}}}
\newcommand{\asrt}{{\textrm{asrt}}}

\newcommand{\cat}[1]{\ensuremath{\textbf{#1}}\xspace}
\newcommand{\Sets}{\ensuremath{\textbf{Sets}}\xspace}
\newcommand{\Top}{\ensuremath{\textbf{Top}}\xspace}
\newcommand{\Meas}{\ensuremath{\textbf{Meas}}\xspace}
\newcommand{\CH}{\ensuremath{\textbf{CH}}\xspace}
\newcommand{\CRng}{\ensuremath{\textbf{CRng}}\xspace}

\newcommand{\PoSets}{\ensuremath{\textbf{PoSets}}\xspace}

\newcommand{\Hilb}{\ensuremath{\textbf{Hilb}}\xspace}

\newcommand{\Vect}{\ensuremath{\textbf{Vect}}\xspace}
\newcommand{\vN}{\ensuremath{{\textbf{W}^*}}\xspace}
\newcommand{\vNkl}{\ensuremath{{\textbf{W}^*_\klbullet}}\xspace}
\newcommand{\vNopkl}{\ensuremath{\op{(\textbf{W}^*_\klbullet)}}\xspace}
\newcommand{\vNop}{\ensuremath{\op{({\textbf{W}^*)}}}\xspace}
\newcommand{\LSub}{\ensuremath{\textbf{LSub}}\xspace}
\newcommand{\CLSub}{\ensuremath{\textbf{CLSub}}\xspace}
\newcommand{\Pred}{\ensuremath{\textsl{Pred}}\xspace}

\newcommand{\Eff}[1]{\ensuremath{[0,1]_{#1}}}

\newcommand{\id}{\ensuremath{\mathrm{id}}}

\newcommand{\klbullet}{{+1}}

\newcommand{\Kl}{\mathcal{K}{\kern-.2ex}\ell}
\newcommand{\EM}{\mathcal{E}{\kern-.2ex}\mathcal{M}}
\newcommand{\DM}{\ensuremath{\mathcal{D}{\kern-.85ex}\mathcal{M}}}
\newcommand{\klafter}{\mathop{\raisebox{.15em}{$\scriptscriptstyle\odot$}}}

\newcommand{\myQEDbox}{\blacksquare}
\newcommand{\myQED}{\hspace*{\fill}$\myQEDbox$}

\newcommand{\uwlim}{\qopname\relax m{uwlim}}

\makeatother
 \newdir{ >}{{}*!/-7.5pt/@{>}}
 \newdir{|>}{!/4.5pt/@{|}*:(1,-.2)@^{>}*:(1,+.2)@_{>}}
 \newdir{ |>}{{}*!/-3pt/@{|}*!/-7.5pt/:(1,-.2)@^{>}*!/-7.5pt/:(1,+.2)@_{>}}
\newcommand{\xyline}[2][]{\ensuremath{\smash{\xymatrix@1#1{#2}}}}
\newcommand{\xyinline}[2][]{\ensuremath{\smash{\xymatrix@1#1{#2}}}^{\rule[8.5pt]{0pt}{0pt}}}
\makeatletter

\makeatletter
\DeclareRobustCommand\widecheck[1]{{\mathpalette\@widecheck{#1}}}
\def\@widecheck#1#2{%
    \setbox\z@\hbox{\m@th$#1#2$}%
    \setbox\tw@\hbox{\m@th$#1%
       \widehat{%
          \vrule\@width\z@\@height\ht\z@
          \vrule\@height\z@\@width\wd\z@}$}%
    \dp\tw@-\ht\z@
    \@tempdima\ht\z@ \advance\@tempdima2\ht\tw@ \divide\@tempdima\thr@@
    \setbox\tw@\hbox{%
       \raise\@tempdima\hbox{\scalebox{1}[-1]{\lower\@tempdima\box
\tw@}}}%
    {\ooalign{\box\tw@ \cr \box\z@}}}
\makeatother
\title{Quotient--Comprehension Chains}
\def\titlerunning{Quotient--Comprehension Chains}

\author{Kenta Cho,
Bart Jacobs,
Bas Westerbaan and
Bram Westerbaan
\institute{Institute for Computing and Information Sciences (iCIS),\\
Radboud University Nijmegen, The Netherlands.}
\email{\{K.Cho,bart,bwesterb,awesterb\}@cs.ru.nl}}
\def\authorrunning{K. Cho, B. Jacobs, B. Westerbaan \& A. Westerbaan}
\def\copyrightholders{K. Cho, B. Jacobs,\\ B. Westerbaan \& A. Westerbaan}

\begin{document}

\maketitle

\begin{abstract}
Quotients and comprehension are fundamental mathematical constructions
that can be described via adjunctions in categorical logic. 
This paper reveals that quotients and comprehension are related to measurement,
not only in quantum logic, but also in probabilistic and classical
logic. This relation is presented by a long series of examples,
some of them easy, and some also highly non-trivial (esp.~for von
Neumann algebras).  We have not yet identified a unifying theory.
Nevertheless, the paper contributes towards such
a theory by introducing the new quotient-and-comprehension perspective
on measurement instruments, and by describing the examples on which
such a theory should be built.
\end{abstract}

\renewcommand{\arraystretch}{1.3}

\section{Introduction}

Measurement is a basic operation in quantum theory: the act of
observing a quantum system. It is characteristic of the quantum world
that such an observation disturbs the system under measurement: it
has a side-effect. In~\cite{Jacobs14} a categorical description of
measurement is given that takes such side-effects into account. We
sketch the essentials, omitting many details. For each predicate $p$
on a type/object $A$ in this theory, there is an `instrument' map
\begin{equation}
\label{diag:instr}
\vcenter{\xymatrix{
A\ar[rr]^-{\instr_{p}} & & A+A
}}
\end{equation}

\noindent that performs the act of measuring $p$. We write $A+A$ for
the coproduct/sum of $A$ with itself, which comes equipped with left
and right insertion/coprojection maps $\kappa_{1}, \kappa_{2} \colon A
\rightarrow A+A$. Intuitively, the map $\instr_{p}$ gives an outcome
in the left summand of $A+A$ if $p$ holds, and in the right
component otherwise. The side-effect associated with the instrument is
the map $\nabla \after \instr_{p} \colon A \rightarrow A$, where
$\nabla = [\idmap,\idmap] \colon A+A \rightarrow A$ is the codiagonal.
If $\nabla \after \instr_{p}$ is the identity map $A \rightarrow A$,
one calls $p$ \emph{side-effect free}. Measurement in a probabilistic
setting is side-effect free, but proper quantum measurement is not.

The set-theoretic case may help to understand this instrument map. For
each predicate $p\subseteq A$ one has $\instr_{p}(a) = \kappa_{1}(a)$
if $a\in P$ and $\instr_{p}(a) = \kappa_{2}(a)$ if $a\not\in
P$. In~\cite{Jacobs14} it is shown that such instrument maps also
exist in a probabilistic and in a quantum setting. In the latter case
one works in the opposite of the category of $C^*$-algebras, with
completely positive unital maps. The instrument~\eqref{diag:instr}
then has type $A\times A \rightarrow A$, and is defined as
$\instr_{p}(a, b) = \sqrt{p}\cdot a \cdot \sqrt{p} + \sqrt{1-p}\cdot b
\cdot \sqrt{1-p}$.  This is the (generalised) L{\"u}ders rule, see for
instance, in~\cite[Eq.(1.3)]{BuschS98}.\footnote{%
Three notions of measurement (instrument) commonly appear
in the literature.
\emph{Sharp} or \emph{projective measurement} corresponds
to~$\instr_p$ where~$p$ is a projection~\cite[\S2.2.5]{NielsenC10},
and appears in von Neumann's projection postulate.
\emph{POVM measurement} corresponds to arbitrary $\instr_p$,
although the post-measurement states are usually left
out~\cite[\S2.2.6]{NielsenC10}.
\emph{Generalized measurements}
capture the different ways the same
POVM can be measured~\cite[\S2.2.3]{NielsenC10};
in the
finite dimensional case, every generalized measurement corresponds
to a composition~$(\varphi+ \psi) \after \instr_p$,
where~$\varphi$ and~$\psi$ are automorphisms.}

The paper~\cite{Jacobs14} lists several requirements for instrument
maps~\eqref{diag:instr}. The question remained: do these requirement
uniquely determine the instrument maps? Put differently: is the
presence of these maps a \emph{property} of a category, or
\emph{structure}? The current paper does not solve this fundamental
problem. But it does uncover the relevance of the logical notions of
quotient and comprehension for measurement.

After the formulation of the theory of instruments~\eqref{diag:instr},
it became clear (see~\cite{Cho15}) that one can also work with \emph{partial} maps $A
\rightarrow A+1$ and $A \rightarrow 1+A$. The two of them can be
combined into a single instrument map $A \rightarrow A+A$ via a
suitable pullback. More importantly, it was noted that in all of the
examples the relevant partial map, called `assert' and written as
$\asrt_{p} \colon A \rightarrow A$ in the category of partial maps, is
a composite of a quotient map $\xi$ and a comprehension map $\pi$, as
in:
\begin{equation}
\label{diag:instrfact}
\vcenter{\xymatrix@R-1.5pc{
A\ar@/_3ex/[dr]_-{\xi}\ar[rr]^-{\asrt_{p}} & & A \\
& \raisebox{-.5em}{$A/p^{\perp} \smash{\;\stackrel{(*)}{=}\;} \set{A}{p}$}
    \ar@/_3ex/[ur]_-{\pi}
}}
\end{equation}

\noindent where $p^{\bot}$ is the negation of $p$. Such a connection
between the fundamental concepts of quotient, comprehension and
measurement is fascinating!  Quotients and comprehension have a clean
description in categorical logic as adjoints (see below for
details). Does that lead to instruments as a property? This question
remains unsolved, but now takes another form:
diagram~\eqref{diag:instrfact} involves an equality, marked with
$(*)$, that seems highly un-categorical: adjoints are determined
up-to-isomorphism, so having an equality between them is
strange. Still this is what we see in all examples, via obvious
choices of quotient and comprehension functors. It is not clear if an
equality (or isomorphism) between a quotient $A/p^{\perp}$ and a
comprehension $\set{A}{p}$ is property or structure. This is a
topic of active research, that requires investigation of
many examples. (We have slightly simplified the
picture~\eqref{diag:instrfact} since there is another operation
$\ceil{p}$ involved, but that is not essential at this stage; it will
be adjusted below.)

This paper is about the following. Once we started looking for
quotients and comprehension in the relevant mathematical models we
found them everywhere, often in somewhat disguised form. Uncovering
familiar constructions, like (co)support for von Neumann algebras, as
quotient and comprehension is mathematically relevant on its own. It
changes one's perspective. Thus, the paper only contains examples.
Many different examples, each showing that certain constructions are
instances of quotient and comprehension. The examples include vector
and Hilbert spaces, sets and topological spaces, various Kleisli
categories of monads used for probability theory, commutative rings,
MV-modules and $C^*$-algebras, and finally (non-commutative) von
Neumann algebras. The examples point to decomposition of
(commutative) mathematical structures as products of quotients and
comprehension, like in ring theory, and used for the sheaf theory of
commutative rings.

In summary, we think that quotients and comprehension provide a new
fruitful perspective on the nature of quantum measurement. This is
illustrated here in many examples. We are fully aware that the
general, final explanation is lacking at this stage. But such a
general theory must be based on a thorough understanding of the
examples. That is the focus of the current paper.

This (missing) underlying general theory will bear some resemblance to
recent work in (non-Abelian) homological algebra, see in
particular~\cite{Weighill14} (where similar adjunction chains are
studied), but also~\cite{Janelidze14,Grandis12}. Part of the
motivation is axiomatising the category of (non-Abelian) groups,
following~\cite{MacLane50}. As a result, stronger properties are used
than occur in the current setting (for instance the first isomorphism
theorem and left adjoints to substitution, corresponding to
bifibrations), which excludes not only our motivating example, the category
of von Neumann algebras,
but also~$\Kl(\mathcal{D})$ and~$\Sets$ to name but two.

\section{Comprehension and Quotients for Vector Spaces}\label{sec:vect}

This section briefly reviews comprehension and quotients for vector
spaces. These constructions are fairly familiar. Their categorical
description via a chain of adjunctions, as in~\eqref{diag:vect} below,
is probably less familiar. This re-description may help to
understand similar such chains in the rest of this paper.

We write $\Vect$ for the category of vector spaces over some fixed
field with linear maps between them.  Linear subspaces are organised
in a category $\LSub$. Its objects are pairs $(V,P)$, where $V$ is a
vector space and $P\subseteq V$ is a linear subspace. A morphism
$(P\subseteq V) \rightarrow (Q\subseteq W)$ in $\LSub$ is a linear map
$f\colon V \rightarrow W$ that restricts to $P \rightarrow Q$, \ie,
that satisfies $P \subseteq f^{-1}(Q)$. There is then an obvious
forgetful functor $\LSub \rightarrow \Vect$. It is a poset
fibration~\cite{Jacobs99}, but that does not play a role here. We view
$\LSub$ as a category of linear predicates, over the category $\Vect$
of linear types.

\parpic[r][l]{%
\begin{minipage}{.5\columnwidth}
\begin{equation}
\label{diag:vect}
\vcenter{\xymatrix{
\LSub\ar[d]_{\dashv\;}^{\;\dashv}
   \ar@/_8ex/[d]^{\;\dashv}_{\begin{array}{c}\scriptstyle\mathrm{Quotient} \\[-.7em]
                        \scriptstyle(P\subseteq V) \mapsto V/P\end{array}}
   \ar@/^8ex/[d]_{\dashv\;}^{\begin{array}{c}\scriptstyle\mathrm{Comprehension} \\[-.7em]
                        \scriptstyle(P\subseteq V) \mapsto P\end{array}} \\
\Vect\ar@/^4ex/[u]^(0.4){0\!}\ar@/_4ex/[u]_(0.4){\!1}
}}
\end{equation}
\end{minipage}
} 

\noindent Interestingly, there is a chain of adjunctions like
in~\eqref{diag:vect}. The up going functors $0, 1 \colon \Vect
\rightarrow \LSub$ are for falsum and truth respectively. They send a
vector space $V$ to the least $0(V) = (\{0\} \subseteq V)$ and
greatest $1(V) = (V \subseteq V)$ subspace. There is a comprehension
functor $(V,P) \mapsto P$ that is right adjoint to truth, and a
quotient functor $(V,P) \mapsto V/P$ that is left adjoint to
falsum. The outer adjunctions involve (natural) bijective
correspondences:
%  I put a new paragraph here

%  to prevent a strange \parpic-bug.
\vspace{-\baselineskip}
$$\begin{prooftree}
\xymatrix{\llap{$1V\! =\,$}(V\!\subseteq\! V)\ar[r]^-{f} & 
   (Q\!\subseteq\! W)} 
\Justifies
\xymatrix{V\ar[r]_-{g} & Q}
\end{prooftree}
\hspace*{5em}
\begin{prooftree}
\xymatrix{(P\!\subseteq\! V)\ar[r]^-{f} & 
   (\{0\} \!\subseteq\! W)\rlap{$\,=\!0W$}} 
\Justifies
\xymatrix{V/P\ar[r]_-{g} & W}
\end{prooftree}$$
\vspace{-\baselineskip}

\noindent The second correspondence says that if $P \subseteq
f^{-1}(\{0\}) = \ker(f)$, then $f$ corresponds to a map $V/P
\rightarrow W$. The quotient uses the equivalence relation $v \sim_{P}
v'$ iff $v-v' \in P$.

The category $\LSub$ is obtained via what is called the `Grothendieck
construction'. Since we will use it many times in the sequel, we make it
explicit. For convenience we restrict it to posets. We write $\PoSets$
for the category of posets with monotone functions between them.

\begin{definition}
\label{def:grothendieck}
Let $\cat{B}$ be a category, with a functor $F\colon\cat{B}
\rightarrow \op{\PoSets}$. We write $\int F$ for the category with
pairs $(X,P)$ as objects, where $X\in\cat{B}$ and $P\in F(X)$. A
morphism $f \colon (X,P) \rightarrow (Y,Q)$ is a map $f\colon X
\rightarrow Y$ in $\cat{B}$ with $P \leq F(f)(Q)$. There is 
an obvious forgetful functor $\int F \rightarrow \cat{B}$, given by
$(X,P) \mapsto X$ and $f \mapsto f$.
\end{definition}

The category $\LSub$ of linear subspaces is obtained via this
Grothendieck construction from the functor $F \colon \Vect
\rightarrow \op{\PoSets}$, where $F(V)$ is the poset of linear
subspaces of $V$, ordered by inclusion; on a linear map $f \colon V
\rightarrow W$ we get $F(f) \colon F(W) \rightarrow F(V)$ by inverse
image: $F(f)(Q) = f^{-1}(Q)$.

The following general observation about the Grothendieck construction
is useful.

\begin{lemma}
\label{lem:grothendieck}
Assume for a functor $F\colon\cat{B} \rightarrow \op{\PoSets}$,
\begin{itemize}
\item each `fibre' $F(X)$ has a least element $0_{X}$;
\item each $F(X)$ also has a greatest element $1_{X}$, and each $F(f)
  \colon F(Y) \rightarrow F(X)$ satisfies $F(f)(1_{Y}) = 1_{X}$.
\end{itemize}

\noindent Then there are functors $0, 1 \colon \cat{B} \rightarrow
\int F$, namely $0(X) = (X,0_{X})$ and $1(X) = (X,1_{X})$, which are
left and right adjoints to the forgetful functor $\int F \rightarrow
\cat{B}$. \myQED
\end{lemma}

\auxproof{
Each $f\colon X \rightarrow Y$ in $\cat{B}$ yields $f \colon (X,0_{X})
\rightarrow (Y, 0_{Y})$ in $\int F$ since $0_{X} \leq
F(f)(0_{Y})$. The adjunction correspondence is trivial:
$$\begin{prooftree}
\xymatrix{(X,0_{X})\ar[r]^-{f} & (Y,Q)} 
\Justifies
\xymatrix{X\ar[r]_-{f} & Y}
\end{prooftree}$$

\noindent since $0_{X} \leq F(f)(Q)$ always holds.

Each $f\colon X \rightarrow Y$ also yields $f \colon (X,1_{X})
\rightarrow (Y, 1_{Y})$ in $\int F$ since $1_{X} \leq
F(f)(1_{Y})$. Here we use that $F(f)$ preserves top elements.
There is then a similar trivial correspondence:
$$\begin{prooftree}
\xymatrix{(X,P)\ar[r]^-{f} & (Y,1_{Y})} 
\Justifies
\xymatrix{X\ar[r]_-{f} & Y}
\end{prooftree}$$

\noindent since $P \leq 1_{X} = F(f)(1_{Y})$ always holds.
}

We briefly sketch the situation for Hilbert spaces, where quotients
are given by (ortho)complements. So let $\Hilb \hookrightarrow \Vect$
be the category of Hilbert spaces, with bounded linear
maps between them. Mapping a Hilbert space $V$ to the poset of closed
linear subspaces yields a functor $\Hilb \rightarrow \op{\PoSets}$. We
write $\CLSub$ for the resulting Grothendieck completion, with
forgetful functor $\CLSub \rightarrow \Hilb$. Since both $\{0\}
\subseteq V$ and $V\subseteq V$ are closed, this functor has both a
left and right adjoint, by Lemma~\ref{lem:grothendieck}.

\parpic[r][l]{%
\begin{minipage}{.5\columnwidth}
\begin{equation}
\label{diag:hilb}
\vcenter{\xymatrix{
\CLSub\ar[d]_{\dashv\;}^{\;\dashv}
   \ar@/_8ex/[d]^{\;\dashv}_{\begin{array}{c}\scriptstyle\mathrm{Quotient} \\[-.7em]
                        \scriptstyle(P\subseteq V) \mapsto P^{\bot}\end{array}}
   \ar@/^8ex/[d]_{\dashv\;}^{\begin{array}{c}\scriptstyle\mathrm{Comprehension} \\[-.7em]
                        \scriptstyle(P\subseteq V) \mapsto P\end{array}} \\
\Hilb\ar@/^4ex/[u]^(0.4){0\!}\ar@/_4ex/[u]_(0.4){\!1}
}}
\end{equation}
\end{minipage}}
We get a situation like in~\eqref{diag:vect}, see~\eqref{diag:hilb}.
For the quotient adjunction, note that if $f \colon V \rightarrow W$
in $\Hilb$ satisfies $P\subseteq \ker(f) = f^{-1}(\{0\})$, for a
closed $P\subseteq V$, then $f$ is determined by its restriction
$P^{\bot} \rightarrow W$, using that $V \cong P \oplus P^{\bot}$. The
latter decomposition of the space $V$ exists because each vector $v\in
V$ can be written in a unique way as $v = v_{1} + v_{2}$ with $v_{1}
\in P$ and $v_{2} \in P^{\bot}$.  This is a basic result in the theory
of Hilbert spaces.

\section{Set-Theoretic Examples}\label{sec:sets}

Standardly it is a \emph{relation} $R\subseteq X\times X$ on a set $X$
that gives rise to a quotient $X/R$, and not a \emph{predicate}, like
for vector spaces in the previous section. Such a quotient $R \mapsto
X/R$ is described as a left adjoint to the equality functor,
see~\cite{Jacobs99} for details. It turns out that a quotient of a
predicate also exists in set-theoretic and other contexts if we switch
to partial functions.  Categorically this will be done via the lift
monad (sometimes called \emph{maybe monad}). We isolate the general
construction first.

\begin{definition}
\label{def:lift}
Let $\cat{B}$ be a category with binary coproducts~$+$ and a final
object~$1$. The functor $X \mapsto X+1$ is then a monad on~$\cat{B}$,
called the lift monad. We write $\cat{B}_{\klbullet}$ for the Kleisli
category of this monad.
\end{definition}

The category $\cat{B}_\klbullet$ thus has the same objects as
$\cat{B}$, and maps $X \rightarrow Y$ in $\cat{B}_\klbullet$ are maps
$X \rightarrow Y+1$ in $\cat{B}$.  We denote the composition in
$\cat{B}_\klbullet$ by $g\klafter f=[g,\kappa_2]\circ f$. For the
category $\Sets$ of sets and functions, the final object is a
singleton~$1=\{*\}$ and coproducts are given by disjoint union.  So
in~$\Sets$ the maps $f\colon X\longrightarrow Y+1\equiv Y\cup \{*\}$
correspond exactly to partial maps from~$X$ to~$Y$.  Hence
$\Sets_\klbullet$ is the category of sets and \emph{partial}
functions.

We define a functor $\square \colon \Sets_{\klbullet} \rightarrow
\op{\PoSets}$ by $\square(X) = \Pow(X)$, the poset of subsets of $X$,
ordered by inclusion. For a function 
$f\colon X \longrightarrow Y+1\equiv Y\cup\{*\}$ we
define $\square(f) \colon \Pow(Y) \rightarrow \Pow(X)$
as:
$$\begin{array}{rcccl}
\square(f)(Q)
& = &
f^{-1}(Q\cup\{*\})
& = &
\setin{x}{X}{\allin{y}{Y}{f(x)=y\implies y\in Q}}.
\end{array}$$

\noindent A morphism~$f\colon (X,P)\longrightarrow (Y,Q)$
in~$\int\square$ is a map $f\colon X\to Y+1$ such that~$f(P)\subseteq
Q\cup\{*\}$.

Each poset $\square(X) = \Pow(X)$ has a greatest element $1 =
X\subseteq X$ and a least element $0 = \emptyset \subseteq
X$. Moreover, $\square(f)(1) = 1$. Hence the conditions of
Lemma~\ref{lem:grothendieck} are satisfied, so that the forgetful
functor $\int\square\rightarrow \Sets_{\klbullet}$ has both a left and
a right adjoint. But there is more.

\begin{proposition}
\label{prop:sets}
In the set-theoretic case we have a chain of adjunctions as shown
in~\eqref{diag:sets} below.
\end{proposition}

\auxproof{
For a map $f \colon 1X \rightarrow (Y,Q)$ in $\int\square$ we have $1
\square_{f}(Q)$, so that $Q(y)$ holds for all $\kappa_{1}y =
f(x)$. Thus $f\colon X \rightarrow X+1$ factors via $X \rightarrow
Q+1$.

For the quotient adjunction we show that there is a bijective
correspondence:
$$\begin{prooftree}
\xymatrix{ (P\subseteq X) \ar[r]^-{f} & 0Y \rlap{$\;=(\emptyset \subseteq Y)$}}
\Justifies
\xymatrix{ \neg P \ar[r]_-{g} & Y }
\end{prooftree}$$

\noindent Indeed, given $f\colon (P\subseteq X) \rightarrow 0Y$ in
$\int\square$, we have $P \subseteq \square_{f}(0) = \set{x}{f(x) = *}$.
Thus $f$ is determined by its outcome on $\neg P$, so that we can
simply define $\overline{f} \colon \neg P \rightarrow Y+1$ as
$\overline{f}(x) = f(x)$. And, in the other direction, for $g \colon
\neg P \rightarrow Y+1$ we define the extension $\overline{g} \colon X
\rightarrow Y+1$ as:
$$\begin{array}{rcl}
\overline{g}(x)
& = &
\left\{\begin{array}{ll}
* \quad & \mbox{if } x\in P \\
g(x) & \mbox{if } x \in \neg P
\end{array}\right.
\end{array}$$

\noindent By construction, this $\overline{g}$ is a map $(P\subseteq
X) \rightarrow (\emptyset \subseteq Y)$ in $\int\square$ since:
$$\begin{array}{rcccl}
\square_{\overline{g}}(0)
& = &
\set{x}{g(x) = *}
& \supseteq &
P.
\end{array}$$

\noindent Then:
$$\begin{array}[b]{rcl}
\overline{\overline{f}}(x)
& = &
\left\{\begin{array}{ll}
* \quad & \mbox{if } x\in P \\
\overline{f}(x) & \mbox{if } x \in \neg P
\end{array}\right. \\
& = &
\left\{\begin{array}{ll}
f(x) \quad & \mbox{if } x\in P \\
f(x) & \mbox{if } x \in \neg P
\end{array}\right. \\
& = &
f(x) \\
\overline{\overline{g}}(x)
& = &
\overline{g}(x) \\
& = &
\left\{\begin{array}{ll}
* \quad & \mbox{if } x\in P \\
g(x) & \mbox{if } x \in \neg P
\end{array}\right. \\
& = &
g(x) \qquad \mbox{since by assumption $x\in\neg P$.}
\end{array}$$
}

\parpic[r][l]{%
\begin{minipage}{.5\columnwidth}
\begin{equation}
\label{diag:sets}
\vcenter{\xymatrix{
\int\square\ar[d]_{\dashv\;}^{\;\dashv}
   \ar@/_8ex/[d]^{\;\dashv}_{\begin{array}{c}\scriptstyle\mathrm{Quotient} \\[-.7em]
                        \scriptstyle(P\subseteq X) \mapsto \neg P\end{array}} 
   \ar@/^8ex/[d]_{\dashv\;}^{\begin{array}{c}\scriptstyle\mathrm{Comprehension} \\[-.7em]
                        \scriptstyle(P\subseteq X) \mapsto P\end{array}} \\
\Sets_{\klbullet}\ar@/^4ex/[u]^(0.4){0\!}\ar@/_4ex/[u]_(0.4){\!1}
}}
\end{equation}
\end{minipage}}
We note that there is a clear similarity with the earlier vector space
and Hilbert space examples: in a quotient $V/P$, for a linear subspace
$P\subseteq V$, all elements from $P$ are identified. Similarly, in
the above set-theoretic case, a subset $P\subseteq X$ yields as
quotient the complement $\neg P = \set{x}{x\not\in P}$. It is the part
of $X$ that remains when all elements from $P$ are removed (or,
identified with the base point, $*$, in a setting with partial
functions). Thus, the quotient of $P\subseteq X$ is the comprehension
of $\neg P$.

The unit of the adjunction between~$0$ and quotient
in~\eqref{diag:sets} for an object~$(X,P)$ in~$\int \square$ is
obtained via the decomposition $X = P + \neg P$. The unit map
$\xi_{P}\colon X\to \neg P+1$ sends~$x\in \neg P$ to itself and~$x\in
P$ to $*\in 1$.  Unfolded the universal property of~$\xi_{P}$ reads:
for every map $f\colon X\to Y+1$ such that~$f(P)\subseteq\{*\}$ there
is unique map~$\overline{f}\colon \neg P\to Y+1$ such
that~$\overline{f}\klafter \xi_{P} = f$.  The map~$\overline{f}$ will
simply be the restriction of~$f$ to~$\neg P$.

The counit of the adjunction between~$1$ and comprehension for an
object~$(Y,Q)$ in~$\int \square$ is the inclusion~$\pi_{Q}\colon Q\to
Y$.  It has the following universal property.  For every~$f\colon
X\to Y+1$ with~$f(X)\subseteq Q \cup \{*\}$ there is a unique map
$\overline{f}\colon X\to Q+1$ such that~$f=\pi_{Q} \klafter
\overline{f}$.  The map~$\overline{f}$ will simply be the restriction
of~$f$ to a partial map from~$X$ to~$Q$.

In this situation consider the following (composite) maps, the first
two in $\Sets_{\klbullet}$, the last one in $\Sets$.
\begin{equation}
\label{diag:setsinstr}
\vcenter{\xymatrix@R-2pc@C-.5pc{
P\ar[r]^-{\pi_{P}} & X\ar[r]^-{\xi_{\neg P}} & P
&
X\ar[r]^-{\xi_{\neg P}} & P\ar[r]^-{\pi_P} & X
&
x\ar[r]^-{\instr_P} & X+X \\
x \ar@{|->}[r] & x \ar@{|->}[r] & x
&
x \ar@{|->}[r] & 
   {\left\{\begin{array}{ll} x & \mbox{if }x\in P \\
   * & \mbox{if } x\not\in P \end{array}\right.} 
\ar@{|->}[r] & 
   {\left\{\begin{array}{ll} x & \mbox{if }x\in P \\
   * & \mbox{if } x\not\in P \end{array}\right.} 
&
x \ar@{|->}[r] & 
   {\left\{\begin{array}{ll} \kappa_{1}x & \mbox{if }x\in P \\
   \kappa_{2}x & \mbox{if } x\not\in P \end{array}\right.} 
}}
\end{equation}

\noindent The first map is the identity; the second one is the
`assert' map $\asrt_P$ from the introduction; and the third one is
obtained by combining $\asrt_P$ and $\asrt_{\neg P}$ via a suitable
pullback. It is the instrument map for measurement, associated with
the predicate $P\subseteq X$.

\parpic[l][l]{%
\begin{minipage}{.5\columnwidth}
\begin{equation}
\label{diag:top}
\vcenter{\xymatrix{
\int\square\ar[d]_{\dashv\;}^{\;\dashv}
   \ar@/_8ex/[d]^{\;\dashv}_{\begin{array}{c}\scriptstyle\mathrm{Quotient} \\[-.7em]
                        \scriptstyle(P\subseteq X) \mapsto \neg P\end{array}} 
   \ar@/^8ex/[d]_{\dashv\;}^{\begin{array}{c}\scriptstyle\mathrm{Comprehension} \\[-.7em]
                        \scriptstyle(P\subseteq X) \mapsto P\end{array}} \\
\Top_{\klbullet}\ar@/^4ex/[u]^(0.4){0\!}\ar@/_4ex/[u]_(0.4){\!1}
}}
\end{equation}
\end{minipage}}
\noindent
There are some relatively straightforward variations of the chain of
adjunctions in~\eqref{diag:sets}.  If one replaces the
poset~$\Pow(X)$ of subsets of a set~$X$ by the poset~$\Clopen(X)$ of
clopens of a topological space~$X$ one gets a functor $\square\colon
\Top\to \op{\PoSets}$.  For a continuous function $f\colon X\to Y+1$
(which corresponds to a continuous partial function $f\colon X\to Y$
with clopen domain) and clopen~$Q\subseteq Y$ we define
$\square(f)(Q)=f^{-1}(Q\cup \{*\})$ as before.  (Note
that~$f^{-1}(Q\cup \{*\})$ is clopen.)  Again one gets a
quotient--comprehension chain~\eqref{diag:top}.  For a
clopen~$P\subseteq X$ the quotient~$\neg P$ and comprehension~$P$ are
the same as in the case of sets~\eqref{diag:sets} but now come with a
natural topology induced by~$X$.  To see that this works one checks
that all maps involved are continuous.

One obtains a similar chain for the category~$\Meas$ of measurable
spaces and measurable maps if one replaces the poset~$\Pow(X)$ of
subsets of a set~$X$ by the poset~$\MeasS(X)$ of measurable subsets of
a measurable space~$X$.

Let us think some more about the chain for topological spaces.  Since
a closed subset of a compact Hausdorff space is again compact, we may
restrict the chain~\eqref{diag:top} to the category~$\CH$ of compact
Hausdorff spaces and the continuous maps between them.  Since~$\CH$ is
dual to a whole slew of `algebraic' categories (as opposed to
`spacial' such as~$\Top$) we get quotient--comprehension chains for
(the opposite of) all those categories as well.  For example, we get a
quotient--comprehension chain for the opposite category of commutative
unital
$C^*$-algebras with unital $*$-homomorphisms 
via Gelfand's duality (see e.g.~\cite{furber13}),
and for
the opposite category of unital Archimedean Riesz spaces with Riesz
homomorphisms via Yosida's duality~\cite{yosida41}.  Interestingly, there are
quotient--comprehension chains for `algebraic' categories which do not
seem to have a `spacial' counterpart such as the category of
commutative rings and homomorphisms, such as the category~$\op{\CRng}$
of commutative rings and homomorphisms, as we will see in
Section~\ref{sec:ring}.

The categories~$\Sets$, $\Top$, $\Meas$, $\CH$ and~$\op{\CRng}$ are
all \emph{extensive}~\cite{Carboni93}.  In fact, any extensive
category~$\mathscr{E}$ with final object has a quotient--adjunction
chain of which~\eqref{diag:sets} and~\eqref{diag:top} are instances.
In particular, any topos will have a quotient--adjunction chain.  In
this general setting, the poset of subsets of a set~$X$ is replaced by
the poset of \emph{complemented} subobjects of an object~$X$
of~$\mathscr{E}$.
Details will appear elsewhere.

For our next example we write $\nePow$ for the \emph{nonempty}
powerset monad on $\Sets$, $\Kl(\nePow)$ for its Kleisli category, and
$\Kl(\nePow)_{\klbullet}$ for the Kleisli category of the lift monad on
$\Kl(\nePow)$. Thus, maps $X \rightarrow Y$ in $\Kl(\nePow)_{\klbullet}$
are functions $X \rightarrow \nePow(Y+1)$. They capture
non-deterministic computation, with multiple successor states and
possibly also non-termination.

There is again a predicate functor $\square \colon
\Kl(\nePow)_{\klbullet} \rightarrow \op{\PoSets}$ with $\square(X) =
\Pow(X)$ for a set~$X$. 
For a map  $f\colon X \rightarrow \nePow(Y+1)$ we define:\quad
\smash{$\begin{array}{rcl}
\square(f)(Q)
& = &
\setin{x}{X}{\allin{y}{Y}{y\in f(x) \Rightarrow Q(y)}}.
\end{array}$}

\begin{proposition}
\label{prop:nepow}
Also for non-deterministic computation via the non-empty
powerset monad $\nePow$ we have
a chain of adjunctions as shown in~\eqref{diag:nepow} below.
\end{proposition}

\begin{myproof}
The truth functor $1(X) = (X\subseteq X)$ and falsum functor $0(X)
= (\emptyset \subseteq X)$ are obtained via
Lemma~\ref{lem:grothendieck}.

\parpic[r][l]{%
\begin{minipage}{.5\columnwidth}
\begin{equation}
\label{diag:nepow}
\vcenter{\xymatrix{
\int\square\ar[d]_{\dashv\;}^{\;\dashv}
   \ar@/_8ex/[d]^{\;\dashv}_{\begin{array}{c}\scriptstyle\mathrm{Quotient} \\[-.7em]
                        \scriptstyle(P\subseteq X) \mapsto \neg P\end{array}} 
   \ar@/^8ex/[d]_{\dashv\;}^{\begin{array}{c}\scriptstyle\mathrm{Comprehension} \\[-.7em]
                        \scriptstyle(P\subseteq X) \mapsto P\end{array}} \\
\Kl(\nePow)_{\klbullet}\ar@/^4ex/[u]^(0.4){0\!}\ar@/_4ex/[u]_(0.4){\!1}
}}
\end{equation}
\end{minipage}}
\noindent
The comprehension adjunction is easy: for a map $f\colon 1X
\rightarrow (Y,Q)$ in $\int\square$, so $f\colon X
\rightarrow \nePow(Y+1)$, we have $1X \subseteq \square(f)(Q)$.
This means that for each $x\in X$ and $y\in Y$ we have: $y\in f(x)
\Rightarrow Q(y)$. Thus we can factor $f$ as $\overline{f} \colon X
\rightarrow \nePow(Q+1)$, giving us a map $\overline{f} \colon X
\rightarrow Q$ in $\Kl(\nePow)_\klbullet$.

\parpic[l][l]{%
\begin{minipage}{.5\columnwidth}
\begin{equation}\label{eq:nepow-quotient}
\hspace{-6em}
\begin{prooftree}
\xymatrix{ (P\subseteq X) \ar[r]^-{f} & 0Y 
   \rlap{\hspace*{2em}in $\int\square$}}
\Justifies
\xymatrix{\neg P \ar[r]_-{g} & Y 
   \rlap{\hspace*{3.5em}in $\Kl(\nePow)_{\klbullet}$}}
\end{prooftree}
\end{equation}
\end{minipage}}
\noindent
The quotient adjunction involves correspondences shown
in~\eqref{eq:nepow-quotient}.  We spell out the transpose operations
of this adjunction below.

Given a map $f\colon (P\subseteq X) \rightarrow (\emptyset \subseteq
Y)$ in $\int\square$, we have $P \subseteq \square(f)(\emptyset) =
\set{x}{* \in f(x)}$. We can define $\overline{f} \colon \neg P
\rightarrow \nePow(Y+1)$ simply as $\overline{f}(x) = f(x)$.

For $g\colon \neg P \rightarrow \nePow(Y+1)$ we get $\overline{g}
\colon X \rightarrow \nePow(Y+1)$ by putting $\overline{g}(x)=g(x)$
for~$x\in P$ and~$\overline{g}(x)=\{*\}$ for~$x\in \neg P$.  This
$\overline{g}$ is a map $(P\subseteq X) \rightarrow (\emptyset
\subseteq Y)$ in $\int\square$ since
$\square(\overline{g})(\emptyset) = \set{x}{\overline{g}(x) = \{*\}}
\supseteq P$.

Then for $x\in X$, we have  $\overline{\overline{f}}(x) = f(x)$, and
also
$\overline{\overline{g}}(x) = g(x)$ for $x\in\neg P$. \myQED

\auxproof{
$$\begin{array}[b]{rcl}
\overline{\overline{f}}(x)
& = &
\left\{\begin{array}{ll}
\{*\} \quad & \mbox{if } x\in P \\
\overline{f}(x) & \mbox{if } x \in \neg P
\end{array}\right. \\
& = &
\left\{\begin{array}{ll}
f(x) \quad & \mbox{if } x\in P \\
f(x) & \mbox{if } x \in \neg P
\end{array}\right. \\
& = &
f(x) \\
\overline{\overline{g}}(x)
& = &
\overline{g}(x) \\
& = &
\left\{\begin{array}{ll}
\{*\} \quad & \mbox{if } x\in P \\
g(x) & \mbox{if } x \in \neg P
\end{array}\right. \\
& = &
g(x) \qquad \mbox{since by assumption $x\in\neg P$.}
\end{array}\eqno{\myQEDbox}$$
}
\end{myproof}

\section{Probabilistic Examples}\label{sec:prob}

In this section we show how the quotient--comprehension chains of
adjunctions also exist in probabilistic computation, via the (finite,
discrete probability) distribution monad $\Dst$ on $\Sets$, and via
the Giry monad $\Giry$ on $\Meas$. The monad $\Dst$ sends a set $X$ to
the set of distributions:
$$\begin{array}{rcl}
\Dst(X)
& = &
\set{r_{1}\ket{x_1} + \cdots r_{n}\ket{x_n}}{r_{i}\in [0,1], x_{i} \in X,
   \sum_{i}r_{i} = 1} \\
& \cong &
\set{\varphi\colon X \rightarrow [0,1]}{\supp(\varphi) \mbox{ is finite, and }
   \sum_{x}\varphi(x) = 1},
\end{array}$$

\noindent where $\supp(\varphi) = \set{x}{\varphi(x) \neq 0}$. The
`ket' notation $\ket{x}$ is just syntactic sugar, used to distinguish
an element $x\in X$ from its occurrence in a formal convex sum in
$\Dst(X)$. In the sequel we shall freely switch between the above two
descriptions of distributions. The unit of the monad is $\eta(x) =
1\ket{x}$, and the multiplication is $\mu(\Phi)(x) =
\sum_{\varphi}\Phi(\varphi) \cdot \varphi(x)$.

We are primarily interested in the Kleisli category $\Kl(\Dst)$ of the
distribution monad. This category has coproducts, like in $\Sets$, and
the singleton set $1 = \{*\}$ as final object, because $\Dst(1) \cong
1$. Hence we can consider the Kleisli category $\Kl(\Dst)_{\klbullet}$
of the lift monad $(-)+1$ on $\Kl(\Dst)$. Its objects are sets, and
its maps $X \rightarrow Y$ are functions $X \rightarrow \Dst(Y+1)$.
Elements of $\Dst(Y+1)$ are  called subdistributions on $Y$.

As before we define a `predicate' functor $\square \colon
\Kl(\Dst)_{\klbullet} \rightarrow \op{\PoSets}$. For a set $X$,
take $\square(X) = [0,1]^{X}$, the set of `fuzzy' predicates $X
\rightarrow [0,1]$ on $X$. They form a poset, by using pointwise the
order on $[0,1]$.  This poset $[0,1]^{X}$ contains a top ($1$) and
bottom ($0$) element, namely the constant functions $x\mapsto 1$ and
$x\mapsto 0$ respectively. For a predicate $p\in [0,1]^{X}$ we write
$p^{\bot} \in [0,1]^{X}$ for the orthocomplement, given by
$p^{\bot}(x) = 1 - p(x)$. Notice that $p^{\bot\bot} = p$, $1^{\bot} =
0$ and $0^{\bot} = 1$. Together with its partial sum operation, the
set of fuzzy predicates $[0,1]^{X}$ forms a what is called an effect
module, that is, an effect algebra with a $[0,1]$-action
(see~\cite{Jacobs14} for details).

A predicate $p\in [0,1]^{X}$ is called \emph{sharp} if $p^{2} = p$.
This means that $p(x) \in \{0,1\}$, so that $p$ is a Boolean predicate
in $\{0,1\}^{X}$. Equivalently, $p$ is sharp if $p \wedge p^{\bot} =
0$. For each predicate $p\in [0,1]^{X}$ there is a least sharp
predicate $\ceil{p}$ with $p \leq \ceil{p}$, and a greatest sharp
predicate $\floor{p} \leq p$, namely:
$$\begin{array}{rclcrcl}
\ceil{p}(x)
& = &
\left\{\begin{array}{ll}
0 \; & \mbox{if } p(x) = 0 \\
1 & \mbox{otherwise.}
\end{array}\right.
& \qquad\qquad &
\floor{p}(x)
& = &
\left\{\begin{array}{ll}
1 \; & \mbox{if } p(x) = 1 \\
0 & \mbox{otherwise.}
\end{array}\right.
\end{array}$$

\noindent It is easy to see that these least sharp and greatest sharp predicates
are each others De Morgan duals, that is, $\floor{p^{\bot}} =
\ceil{p}^{\bot}$. If $p$ is a sharp, then $\floor{p} = p =
\ceil{p}$.

For a function $f\colon X \rightarrow \Dst(Y+1)$ we define $\square(f)
\colon [0,1]^{Y} \rightarrow [0,1]^{X}$ as:
$$\begin{array}{rcl}
\square(f)(q)(x)
& = &
\sum_{y\in Y} f(x)(y) \cdot q(y) + f(x)(*).
\end{array}$$

\noindent Since $f(x) \in \Dst(Y+1)$ is a distribution, we have
$\sum_{y\in Y} f(x)(y) + f(x)(*) = 1$, so that $\square(f)(1) =
1$. Hence Lemma~\ref{lem:grothendieck} applies, so that we have a
functor $\int\square \rightarrow \Kl(\Dst)_{\klbullet}$ with falsum $0$
as left adjoint, and truth $1$ as right adjoint. Recall that a map
$(X,p) \rightarrow (Y,q)$ in $\int\square$ is a function $f\colon X
\rightarrow \Dst(Y+1)$ with $p(x) \leq \square(f)(q)(x)$ for all
$x\in X$.

\parpic[r][l]{%
\begin{minipage}{.5\columnwidth}
\vspace{-1ex}
\begin{equation}
\label{diag:dst}
\vcenter{\xymatrix{
\int\square\ar[d]_{\dashv\;}^{\;\dashv}
   \ar@/_8ex/[d]^{\;\dashv}_{\begin{array}{c}\scriptstyle\mathrm{Quotient} \\[-.7em]
                        \scriptstyle(p\in[0,1]^{X}) \mapsto X/p\end{array}} 
   \ar@/^8ex/[d]_{\dashv\;}^{\begin{array}{c}\scriptstyle\mathrm{Comprehension} \\[-.7em]
                        \scriptstyle(p\in[0,1]^{X}) \mapsto \cmpr{X}{p}\end{array}} \\
\Kl(\Dst)_{\klbullet}\ar@/^4ex/[u]^(0.4){0\!}\ar@/_4ex/[u]_(0.4){\!1}
}}
\end{equation}
\end{minipage}}
\begin{proposition}
\label{prop:dist}
The distribution monad $\Dst$ on $\Sets$, used to model probabilistic
computation, gives rise to the chain of adjunctions~\eqref{diag:dst}
to the right where $\cmpr{X}{p} = \setin{x}{X}{p(x) = 1}$, and $X/p =
\cmpr{X}{\ceil{p^{\bot}}} = \set{x}{p(x)\neq 1}$.
\end{proposition}

\vspace{0ex}
\begin{myproof}
For a map $f\colon 1Y \rightarrow (X,p)$ in $\int\square$ we have
$f\colon Y \rightarrow \Dst(X+1)$ satisfying $1 \leq \square(f)(p)$.
This means $1 = \big(\sum_{x} f(y)(x) \cdot p(x)\big) + f(y)(*)$, for
each $y\in Y$. Since $\sum_{x}f(y)(x) + f(y)(*) = 1$, this can only
happen if $f(y)(x) \neq 0 \Rightarrow p(x) = 1$. But then we can
factor $f$ as $\overline{f} \colon Y \rightarrow \cmpr{X}{p}$ in
$\Kl(\Dst)_{\klbullet}$, where $\overline{f}(y) = \sum_{x, f(y)(x)\neq
  0} f(y)(x)\ket{x} + f(y)(*)\ket{*}$.

In the other direction, given a function $g\colon Y \rightarrow
\Dst(\cmpr{X}{p}+1)$ we define the map $\overline{g} \colon Y \rightarrow
\Dst(X+1)$ as $\overline{g}(y) = \sum_{x, p(x)=1} g(y)(x)\ket{x} +
g(y)(*)\ket{*}$. Then, for each $y\in Y$,
$$\begin{array}{rcl}
\square(\overline{g})(p)(y)
& = &
\sum_{x, p(x) = 1} \overline{g}(y)(x)\cdot p(x) + \overline{g}(y)(*) 
 = 
\sum_{x, p(x) = 1} g(x)(y) + g(y)(*)
 = 
1.
\end{array}$$

\auxproof{
\noindent Moreover, for $y\in Y$,
$$\begin{array}{rcl}
\overline{\overline{f}}(y)
& = &
\sum_{x, p(x)=1} \overline{f}(y)(x)\ket{x} + \overline{f}(y)(*)\ket{*} 
\hspace*{\arraycolsep}=\hspace*{\arraycolsep}
\sum_{x} f(y)(x)\ket{x} + \overline{f}(y)(*)\ket{*} 
\hspace*{\arraycolsep}=\hspace*{\arraycolsep}
f(y) \\
\overline{\overline{g}}(y)
 &=& 
\sum_{x, \overline{g}(y)(x)\neq 0} \overline{g}(y)(x)\ket{x} 
   + \overline{g}(y)(*)\ket{*} 
\hspace*{\arraycolsep}=\hspace*{\arraycolsep}
\sum_{x\in\cmpr{X}{p}} g(y)(x)\ket{x} + g(y)(*)\ket{*} 
\hspace*{\arraycolsep}=\hspace*{\arraycolsep}
g(y).
\end{array}$$
}

\parpic[r][l]{%
\begin{minipage}{.5\columnwidth}
\begin{equation}
\label{eq:prob-corr}
\begin{prooftree}
\xymatrix{ (X,p) \ar[r]^-{f} & 0Y}
\Justifies
\xymatrix{ X/p \ar[r]_-{g} & Y }
\end{prooftree}
\end{equation}
\end{minipage}
}
The quotient adjunction involves the correspondence~\eqref{eq:prob-corr},
which works as follows. Given $f\colon (X,p) \rightarrow 0Y$
in $\int\square$, then $f\colon X \rightarrow \Dst(Y+1)$ satisfies $p
\leq \square(f)(0)$. This means that $p(x) \leq \sum_{y} f(x)(y)\cdot
0(y) + f(x)(*) = f(x)(*)$, for each $x\in X$. We then define
$\overline{f} \colon X/p \rightarrow \Dst(Y+1)$ as $\overline{f}(x) =
\sum_{y} \frac{f(x)(y)}{p^{\bot}(x)}\ket{y} + \frac{f(x)(*) -
  p(x)}{p^{\bot}(x)}\ket{*}$. This is well-defined, since $p^{\bot}(x)
\neq 0$ for $x\in X/p$.

\auxproof{
Moreover, $\overline{f}(x)$ is a distribution since:
$$\begin{array}{rcccccl}
\big(\sum_{y} \frac{f(x)(y)}{p^{\bot}(x)}\big) + 
   \frac{f(x)(*) - p(x)}{p^{\bot}(x)}
& = &
\frac{1-f(x)(*) + f(x)(*) - p(x)}{p^{\bot}(x)}
& = &
\frac{1- p(x)}{p^{\bot}(x)}
& = &
1.
\end{array}$$

\noindent Notice that if $f$ is Cartesian, that is, if $\square(f)(0)
= p$, then $f(x)(*) = p(x)$, so that $\overline{f}(x)(*) =
\frac{f(x)(*) - p(x)}{p^{\bot}(x)} = 0$. Thus, $f$ is a total map.
}

In the other direction, given $g \colon X/p \rightarrow \Dst(Y+1)$ we
define $\overline{g} \colon X \rightarrow \Dst(Y+1)$ as:
$$\begin{array}{rcl}
\overline{g}(x)
& = &
\sum_{y} p^{\bot}(x)\cdot g(x)(y)\ket{y} 
 \; + \;
   \big(p(x) + p^{\bot}(x)\cdot g(x)(*)\big)\ket{*}.
\end{array}$$

\noindent Notice that this extension of $g$ outside the subset
$\cmpr{X}{\ceil{p^{\bot}}} \hookrightarrow X$ is well-defined, since
if $x \not\in \cmpr{X}{\ceil{p^{\bot}}}$, then $p(x) = 1$, so
$p^{\bot}(x) = 0$, which justifies writing $p^{\bot}(x)\cdot
g(x)(y)$. In that case, when $p(x)=1$, we get $\overline{g}(x) =
1\ket{*}$. This $\overline{g}$ is a morphism $(X,p) \rightarrow 0Y$ in
$\int\square$, since $p \leq \square(\overline{g})(0)$, that is $p(x)
\leq \overline{g}(x)(*)$. This follows since $p^{\bot}(x) \geq 0$ and
$g(x)(*) \geq 0$ in $\overline{g}(x)(*) = p(x) + p^{\bot}(x)\cdot
g(x)(*) \geq p(x)$. \myQED

\auxproof{
Moreover, $\overline{g}(x)$ is a well-defined distribution in
$\Dst(Y+1)$ since:
$$\begin{array}{rcl}
\lefteqn{\textstyle\big(\sum_{y} p^{\bot}(x)\cdot g(x)(y)\big) + 
   \big(p(x) + p^{\bot}(x)\cdot g(x)(*)\big)} \\
& = &
p^{\bot}(x)\cdot (1-g(x)(*)) + p(x) + p^{\bot}(x)\cdot g(x)(*) \\
& = &
p^{\bot}(x) + p(x) \\
& = &
1.
\end{array}$$
}

\auxproof{
Notice that if $g$ is a total map, that is $g(x)(*) = 0$ for
each $x$, then $\square_{\overline{g}}(0)(x) = p(x) + p^{\bot}(x)\cdot
g(x)(*) = p(x)$.  Thus $\square_{\overline{g}}(0) = p$, making
$\overline{g}$ Cartesian.
}

\auxproof{
\noindent Finally we have to prove that the mappings $f \mapsto \overline{f}$
and $g \mapsto \overline{g}$ are each others inverses.
$$\begin{array}[b]{rcl}
\overline{\overline{f}}(x)
& = &
\sum_{y} p^{\bot}(x)\cdot \overline{f}(x)(y)\ket{y} + 
   \big(p(x) + p^{\bot}(x)\cdot \overline{f}(x)(*)\big)\ket{*} \\
& = &
\sum_{y} f(x)(y)\ket{y} + \big(p(x) + f(x)(*) - p(x)\big)\ket{*} \\
& = &
\sum_{y} f(x)(y)\ket{y} + f(x)(*)\ket{*} \\
& = &
f(x) \\
\overline{\overline{g}}(x)
& = &
\sum_{y} \frac{\overline{g}(x)(y)}{p^{\bot}(x)}\ket{y} + 
   \frac{\overline{g}(x)(*) - p(x)}{p^{\bot}(x)}\ket{*} \\
& = &
\sum_{y} g(x)(y)\ket{y} + \frac{p^{\bot}(x)\cdot g(x)(*)}{p^{\bot}(x)}\ket{*} \\
& = &
\sum_{y} g(x)(y)\ket{y} + g(x)(*)\ket{*} \\
& = &
g(x).
\end{array}$$
}
\end{myproof}

The counit map $\pi_{p} \colon \cmpr{X}{p} \rightarrow \Dst(X+1)$ and
the unit $\xi_{p} \colon X \rightarrow \Dst(X/p + 1)$ are given by
$\pi_{p}(x) = 1\ket{x}$ and $\xi_{p}(x) = p^{\bot}(x)\ket{x} +
p(x)\ket{\!*\!}$. We can consider their combination, like in
diagram~\eqref{diag:setsinstr}, in $\Kl(\Dst)_{\klbullet}$.
%\begin{equation}
%\label{diag:dstinstr}
$$\vcenter{\xymatrix@R-2pc@C-1pc{
\cmpr{X}{\ceil{p}}\ar[r]^-{\pi_{\ceil{p}}} & X\ar[r]^-{\xi_{p^{\bot}}} & 
   X/p^{\bot} = \cmpr{X}{\ceil{p}}
&
X\ar[r]^-{\xi_{p^{\bot}}} & X/p^{\bot} = \cmpr{X}{\ceil{p}}\ar[r]^-{\pi_{\ceil{p}}} & X
\\
x \ar@{|->}[r] & 1\ket{x} \ar@{|->}[r] & p(x)\ket{x} + p^{\bot}(x)\ket{*}
&
x \ar@{|->}[r] & p(x)\ket{x} + p^{\bot}(x)\ket{*} \ar@{|->}[r] & 
   p(x)\ket{x} + p^{\bot}(x)\ket{*}
}}$$
%\end{equation}

\noindent The map on the left is the identity if the predicate $p$ is
sharp. The map on the right is the `assert' map $\asrt_p$, which
yields, together with $\asrt_{p^{\bot}}$ the instrument $\instr_{p}
\colon X \rightarrow X+X$ in $\Kl(\Dst)$ given by $\instr_{p}(x) =
p(x)\ket{\kappa_{1}x} + (1-p(x))\ket{\kappa_{2}x}$, precisely as
in~\cite{Jacobs14}.

%% For the quotient $X/p^{\bot} = \cmpr{X}{\ceil{p}} =
%% \set{x}{p(x) \neq 0}$ we can form in $\Kl(\Dst)_{\klbullet}$, the
%% composites:
%% $$\xymatrix@R-.5pc{
%% X\ar[r]^-{\xi_{p^\bot}}\ar[dr]_{\asrt_p} & 
%%    X/p^{\bot}\ar[d]^{\pi_{\ceil{p}}} 
%% &
%% &
%% X/p^{\bot}\ar[dr]\ar[r]^-{\pi_{\ceil{p}}} & X\ar[d]^{\xi_{p^{\bot}}}
%% \\
%% & X
%% &
%% &
%% & X/p^{\bot}
%% }$$

%% \noindent On the left the `assert $p$' function $\asrt_{p} \colon X
%% \rightarrow \Dst(X+1)$ is:
%% $$\begin{array}{rcccccl}
%% \asrt_{p}(x)
%% & = &
%% \big(\pi_{\ceil{p}} \klafter \xi_{p^\bot}\big)(x)
%% & = &
%% p(x)\ket{x} + p^{\bot}(x)\ket{*}.
%% \end{array}$$

%% \noindent This map forms the basis for (measuring) instruments (and is
%% essentially as in~\cite{Jacobs14}).  The other composite yields, for
%% $x\in X/p^{\bot}$ and thus $p(x) \neq 0$,
%% \quad\smash{$\begin{array}{rcl}
%% \big(\xi_{p^\bot} \klafter \pi_{\ceil{p}}\big)(x)
%% & = &
%% p(x)\ket{x} + p^{\bot}(x)\ket{*}.
%% \end{array}$}

%% \noindent Hence this composite is the identity if $p$ is sharp.

We can generalise the situation from (finite) discrete probabilistic
computation via the monad $\Dst$, to continuous probabilistic
computation via the Giry monad $\Giry$ on the category $\Meas$ of
measurable spaces and measurable functions. The category
$\Kl(\Giry)_\klbullet$ of partial maps in the associated Kleisli
category is isomorphic to the Kleisli category $\Kl(\Gsub)$ of the
`subprobability' Giry monad. We prefer to work with the latter. Thus,
for a measurable space $(X,\Sigma_X)$, which is referred to simply by
$X$, we set:
$$\begin{array}{rcl}
\Gsub(X)
& = &
\set{\phi\colon \Sigma_X\to [0,1]}{\text{$\phi$ is a subprobability measure}},
\end{array}$$

\noindent where a \emph{subprobability measure} is a countably
additive map $\phi\colon \Sigma_X\to [0,1]$ with $\phi(\emptyset)=0$,
but not necessarily $\phi(X)=1$. As predicates $\Pred(X)$ on $X\in\Meas$
we use measurable functions $X \rightarrow [0,1]$. They form an
effect module, see~\cite{Jacobs13} for details.

\auxproof{%
The Kleisli category $\Kl(\Giry)$
inherits coproducts from $\Meas$.
The binary coproduct $X+Y$ of measurable spaces
is the disjoint union of the underlying sets
with the $\sigma$-algebra $\Sigma_{X+Y}$
consisting of subsets of the form $M+N$
for $M\in\Sigma_X$ and $N\in\Sigma_Y$.
Namely one has $\Sigma_{X+Y}\cong\Sigma_X\times\Sigma_Y$.
The singleton measurable space $1$
is final in $\Meas$, so is in $\Kl(\Giry)$
because $\Giry(1)\cong 1$.
Like before,
we will use the Kleisli category $\Kl(\Giry)_\klbullet$
of the lift monad on $\Kl(\Giry)$.
Maps $f\colon X\to Y$ in $\Kl(\Giry)_\klbullet$
are measurable functions $f\colon X\to \Giry(Y+1)$.
The set $\Giry(Y+1)$ is identified with the set of
subprobability measures $\phi\colon \Sigma_Y\to[0,1]$,
and hence maps in $\Kl(\Giry)_\klbullet$
are substochastic relations.
}%
\auxproof{%
We have a bijection $\alpha\colon\Giry(X+1)\cong\Gsub(X)$:
for $\phi\in\Giry(X+1)$
define $\alpha(\phi)(M)\in\Gsub(X)$
by $\alpha(\phi)=\phi(M+\emptyset)$;
and for $\psi\in\Gsub(X)$
define $\alpha^{-1}(\psi)\in\Giry(X+1)$
by $\alpha^{-1}(\psi)(M+\emptyset)=\psi(M)$
and $\alpha^{-1}(\psi)(M+1)=\psi(M)+(1-\psi(X))$.
Writing $\ev'_M\colon\Gsub(X)\to[0,1]$
for evaluation maps for $\Gsub$,
we have $(\ev'_{M}\circ \alpha)(\phi)=
\alpha(\phi)(M)=\phi(M+\emptyset)=\ev_{M+\emptyset}(\phi)$;
and $(\ev_{M+\emptyset}\circ\alpha^{-1})(\psi)
=\alpha^{-1}(\psi)(M+\emptyset)=\psi(M)=\ev'_M(\psi)$
and $(\ev_{M+1}\circ\alpha^{-1})(\psi)
=\alpha^{-1}(\psi)(M+1)=\psi(M)+(1-\psi(X))=\ev'_M(\psi)
+(1-\psi(X))$. It follows that
both $\alpha$ and $\alpha^{-1}$ are measurable.
}%
\auxproof{%
For each $M\in\Sigma_X$
we have an `evaluation' map $\ev_M\colon\Gsub(X)\to[0,1]$
given by $\ev_M(\phi)=\phi(M)$.
We equip $\Gsub(X)$ with the least $\sigma$-algebra
making $\ev_M$ measurable for all $M\in\Sigma_X$.
Note that for $X,Y\in\Meas$,
a function $f\colon X\to \Gsub(Y)$
is measurable if and only if
$\ev_N\circ f=f(-)(N)\colon X\to[0,1]$ is measurable for each $N\in\Sigma_Y$.
\auxproof{(if)
The $\sigma$-algebra $\Sigma_{\Gsub Y}$ is generated by
$(\ev_N)^{-1}(A)$ for $N\in \Sigma_Y$ and $A\in\Sigma_{[0,1]}$,
while we have $f^{-1}((\ev_N)^{-1}(A))=(ev_N\circ f)^{-1}(A)\in\Sigma_X$.
}%
For a measurable function $f\colon X\to Y$,
define $\Gsub(f)\colon\Gsub(X)\to\Gsub(Y)$
by $\Gsub(f)(\phi)=\phi\circ f^{-1}$,
where $f^{-1}\colon \Sigma_Y\to\Sigma_X$
is the inverse image map.
The functor $\Gsub\colon\Meas\to\Meas$ is
a monad via the following data.
The unit $\eta_X\colon X\to\Gsub(X)$ is given by
$\eta_X(x)(M)=\indic{M}(x)$, namely $\eta_X(x)$
is the Dirac measure at $x$.
For a map $f\colon X\to\Gsub(Y)$ in $\Meas$,
the Kleisli extension $f_*\colon\Gsub(X)\to\Gsub(Y)$
is given by $f_*(\phi)(N)=\int f(-)(N)\,d\phi$.
}%

Now we define a predicate functor $\square \colon \Kl(\Gsub)
\rightarrow \op{\PoSets}$.  For a measurable space $X$ we define
$\square(X)=\Pred(X)$.  For a Kleisli map $f\colon X\to \Gsub(Y)$,
define $\square(f)\colon\Pred(Y)\to\Pred(X)$ by integration:
$\square(f)(q)(x) = \int q\intd f(x) \,+\, (1-f(x)(Y))$.

\auxproof{%
We check that $\square(f)(q)\colon X\to[0,1]$
is measurable.
Note that $f(-)(Y)$ is measurable by definition.
Assume $q=\lim_{n\to\infty} q_n$
for step functions $q_n$,
so that $\int_Y q\,d f(x)=\lim_{n\to\infty}
\int_Y q_n\,d f(x)$,
see~\cite[Lemma~2 and Definition~3]{Jacobs13}.
Then it is not hard to see $x\mapsto \int_Y q_n\,d f(x)$
is measurable for each $n$.
Therefore $x\mapsto \int_Y q\,d f(x)$
is a pointwise limit of measurable functions,
which is also measurable.
Note that
$\square(f)(1)(x)
=\int_Y 1\,d f(x) +(1-f(x)(Y))
=1$,
and hence $\square(f)(1) =
1$. By Lemma~\ref{lem:grothendieck},
we have a functor $\int\square \rightarrow \Kl(\Gsub)$
with falsum $0$
as left adjoint, and truth $1$ as right adjoint.
}%

\begin{proposition}
For the `subprobability' Giry monad $\Gsub$ on $\Meas$ there is the
chain of adjunctions~\eqref{diag:giry} where $\cmpr{X}{p} =
\setin{x}{X}{p(x) = 1}$, and $X/p = \set{x}{p(x)\neq 1}$.
\end{proposition}
\parpic[r][l]{%
\begin{minipage}{.5\columnwidth}
\begin{equation}
\label{diag:giry}
\vcenter{\xymatrix{
\int\square\ar[d]_{\dashv\;}^{\;\dashv}
   \ar@/_8ex/[d]^{\;\dashv}_{\begin{array}{c}\scriptstyle\mathrm{Quotient} \\[-.7em]
                        \scriptstyle p \mapsto X/p\end{array}} 
   \ar@/^8ex/[d]_{\dashv\;}^{\begin{array}{c}\scriptstyle\mathrm{Comprehension} \\[-.7em]
                        \scriptstyle p \mapsto \cmpr{X}{p}\end{array}} \\
\Kl(\Gsub)\ar@/^4ex/[u]^(0.4){0\!}\ar@/_4ex/[u]_(0.4){\!1}
}}
\end{equation}
\end{minipage}}
\begin{myproof}
The verification's proceed much like for Proposition~\ref{prop:dist},
with summation $\sum$ for discrete distributions replaced by
integration $\int$ for continuous distributions. Details are left to
the interested reader. \myQED

\auxproof{%
Note first that
$\cmpr{X}{p}=\setin{x}{X}{p(x) = 1}=p^{-1}(\{1\})\in\Sigma_X$.
Hence the set $\cmpr{X}{p}$ inherits a $\sigma$-algebra
from $X$. The inclusion $\cmpr{X}{p}\hookrightarrow X$
is in $\Meas$, so we have a map
$\pi_p\colon \cmpr{X}{p}\to\Gsub(X)$
by composing the unit of $\Gsub$, \ie,
$\pi_p(x)=\eta_X(x)$.
In fact, this is a map $\pi_p\colon1\cmpr{X}{p}\to
(X,p)$ in $\int\square$,
which satisfies $1\le\square(\pi_p)(p)$,
since for all $x\in\cmpr{X}{p}$ one has
\[
\square(\pi_p)(p)(x)
=\int_X p \,d \pi_p(x)+
(1-\pi_p(x)(X))
=\int_X p \,d \eta_X(x)=p(x)=1
\enspace.
\]
We claim that $\pi_p$ gives a counit
for the rightmost adjunction.
Let $f\colon 1Y \rightarrow (X,p)$
be a map in $\int\square$, \ie,
a measurable function
$f\colon Y \rightarrow \Gsub(X)$ satisfying $1 \leq
\square(f)(p)$.  This means, for each $y\in Y$,
$1 = \int_X p\,d f(y) + (1-f(y)(X))$
\ie, $\int_X p\,d f(y)=f(y)(X)$.
Note that $\int_{\cmpr{X}{p}} p\,d f(y)=
\int_{\cmpr{X}{p}} 1\,d f(y)
=f(y)(\cmpr{X}{p})$
and hence we obtain
\[
\int_{X\setminus\cmpr{X}{p}} p^\bot\,d f(y)
=\int_{X\setminus\cmpr{X}{p}} 1\,d f(y)-
\int_{X\setminus\cmpr{X}{p}} p\,d f(y)
=f(y)(X)-\int_X p\,d f(y)=0
\]
Since $p^\bot(x)>0$ for all $x\in X\setminus\cmpr{X}{p}$,
it follows that $f(y)(X\setminus\cmpr{X}{p})=0$.
\auxproof{%
Let $M=X\setminus\cmpr{X}{p}$ and
$M_n = \{x\in M\mid 1/(n+1)<p^\bot(x)\le 1/n\}$.
Then
$f(y)(M_n)
=\int_{M} \indic{M_n} \,d f(y)
\le \int_{M} (n+1) p^{\bot}\,d f(y)
= (n+1) \int_{M} p^{\bot}\,d f(y)=0$.
Now $M=\biguplus_{n=1}^\infty M_n$
and hence $f(y)(M)=\sum_{n=1}^\infty f(y)(M_n)=0$.
}%

Now, define $\overline{f} \colon Y \rightarrow \cmpr{X}{p}$ in
$\Kl(\Gsub)$ by
$\overline{f}(y)(M)=f(y)(M)$.
To see $\pi_p\klafter\overline{f}=f$,
first note that the Kleisli extension
$(\pi_p)_*\colon\Gsub(\cmpr{X}{p})\to\Gsub(X)$
is given by
\[
(\pi_p)_*(\phi)(M)=
\int_{\cmpr{X}{p}} \pi_p(-)(M)\,d\phi
=\int_{\cmpr{X}{p}} \indic{M\cap\cmpr{X}{p}}\,d\phi
=\phi(M\cap\cmpr{X}{p})
\enspace.
\]
Then, for $y\in Y$ and $M\in\Sigma_X$,
\begin{align*}
f(y)(M)
&= f(y)(M\cap\cmpr{X}{p}) +
f(y)(M\cap(X\setminus\cmpr{X}{p})) \\
&= f(y)(M\cap\cmpr{X}{p})
&&\text{since $f(y)(X\setminus\cmpr{X}{p})=0$} \\
&=\overline{f}(y)(M\cap\cmpr{X}{p}) \\
&= ((\pi_p)_*\circ\overline{f})(y)(M) \\
&= (\pi_p\klafter\overline{f})(y)(M)
\enspace.
\end{align*}
Therefore $\pi_p\klafter\overline{f}=f$.
The uniqueness of $\overline{f}$ follows from
the injectivity of $(\pi_p)_*$.
We have shown there is a `comprehension' right adjoint
to the truth functor $1\colon \Kl(\Gsub)\to\int\square$.

Next, we show that the quotient $X/p$ gives a left adjoint
to the falsum functor $1\colon \Kl(\Gsub)\to\int\square$.
Note that $X/p=\{x\in X\mid p(x)\ne 1\}=p^{-1}([0,1))\in\Sigma_X$
and hence $X/p$ inherits a $\sigma$-algebra from $X$.
For $(X,p)\in\int\square$,
we define a unit $\xi_p\colon (X,p)\to 0(X/p)$
of the adjunction by $\xi_p(x)(M)=p^\bot(x)\indic{M}(x)$.
This indeed defines a measurable function $\xi_p\colon X\to \Gsub(X/p)$,
since for each $M\in\Sigma_{X/p}$
the composite $\ev_M\circ\xi_p=p^\bot\cdot\indic{M}$
is a pointwise product of measurable functions
$p^\bot$ and $\indic{M}$. Moreover it satisfies
$p\le \square(\xi_p)(0)$: for each $x\in X$,
\[
\square(\xi_p)(0)(x)=1-\xi_p(x)(X/p)
=1-p^\bot(x)\indic{X/p}(x)
\ge 1-p^\bot(x)=p(x)
\enspace.
\]
Now we show that these maps $\xi_p$ have the universality
as a unit of adjunction.
Let $f\colon (X,p)\to 0Y$ be a map in $\int\square$,
that is, a measurable function $f\colon X\to\Gsub(Y)$ with
$p\le \square(f)(0)$. The inequality means
$p(x)\le 1-f(x)(Y)$, \ie, $f(x)(Y)\le p^\bot(x)$
for each $x\in X$.
Then, define $\overline{f}\colon X/p\to Y$ in $\Kl(\Gsub)$ by
$\overline{f}(x)(N)=f(x)(N)/p^\bot(x)$.
Note that $p^\bot(x)>0$ for all $x\in X/p$,
and $\overline{f}(x)$ is a subprobability measure on $Y$ since
$f(x)(Y)/p^\bot(x)\le 1$.
Note also that $\overline{f}$ is measurable, since
for each $N\in\Sigma_Y$, the composite $\ev_N\circ\overline{f}=
f(-)(N)/p^\bot$ is a pointwise quotient of measurable functions.
We wish to show $\overline{f}\klafter\xi_p=f$,
\ie, $(\overline{f}\klafter\xi_p)(x)(N)
\coloneqq \int_{X/p} \overline{f}(-)(N)\,d\xi_p(x)
= f(x)(N)$ for $x\in X$ and $N\in\Sigma_Y$.
We will prove this by cases.
\begin{enumerate}
\item\label{enum:x-in-quotient} (When $x\in X/p$)
We have $\xi_p(x)=p^\bot(x)\eta_{X/p}(x)$.
Then
\begin{align*}
\int_{X/p} \overline{f}(-)(N)\,d\xi_p(x)
&=\int_{X/p} \overline{f}(-)(N)\,d(p^\bot(x)\eta_{X/p}(x)) \\
&=p^\bot(x)\int_{X/p} \overline{f}(-)(N)\,d\eta_{X/p}(x) \\
&=p^\bot(x)\overline{f}(x)(N) \\
&=p^\bot(x)(f(x)(N)/p^\bot(x)) \\
&=f(x)(N)
\end{align*}
\item (When $x\in X\setminus(X/p)$)
We have $p^\bot(x)=0$
and hence $\xi_p(x)=0$.
Then $\int_{X/p} \overline{f}(-)(N)\,d\xi_p(x)=0$.
Now, recall that $f$ satisfies $f(x)(Y)\le p^\bot(x)$.
Therefore $f(x)(Y)=0$ and thus $f(x)(N)=0$.
\end{enumerate}
Finally, to see the uniqueness of $\overline{f}$,
let $h,k\colon X/p\to Y$ be maps in $\Kl(\Gsub)$
and assume $h\klafter \xi_p=k\klafter\xi_p$.
For $x\in X/p$ and $N\in\Sigma_Y$,
by the same reasoning as the case~(\ref{enum:x-in-quotient})
above,
\begin{align*}
p^\bot(x)h(x)(N)
=\int_{X/p} h(-)(N)\,d\xi_p(x)
&=(h\klafter \xi_p)(x)(N) \\
&=(k\klafter\xi_p)(x)(N)
=\int_{X/p} k(-)(N)\,d\xi_p(x)
=p^\bot(x)h(x)(N)
\enspace.
\end{align*}
Then $h(x)(N)=k(x)(N)$ because $p^\bot(x)>0$.
Hence $h=k$.
\myQED
}%
\end{myproof}

\section{Commutative Ring Examples}\label{sec:ring}

In Section~\ref{sec:sets} it was mentioned that extensive categories
have quotient--comprehension chains. This applies in particular to the
extensive category $\op{\CRng}$, where $\CRng$ is the category of
commutative rings. Nevertheless, we describe the ring-theoretic
construction here in some detail, because (1)~it forms a good
preparation for the more complicated example of von Neumann algebras in
the next section, (2)~it points to a relation with decomposition in
the sheaf theory of rings.

An element $e\in R$ in a ring is called idempotent if $e^{2} = e$.
The set $\Pred(R)$ of idempotents in $R$ is an effect algebra in
general, and a Boolean algebra if $R$ is commutative. We concentrate
on the latter case; then $e \leq d$ iff $ed = e$, with $e\wedge d =
ed$ and $e^{\bot} = 1-e$. The ring of integers $\Z$ is initial in
$\CRng$, and thus final in $\op{\CRng}$. The Kleisli category
$\op{\CRng}_{\klbullet}$ of the lift monad has ring homomorphisms
$R\times\mathbb{Z} \rightarrow S$ as maps $S \rightarrow R$. They
correspond to \emph{subunital} maps $R \rightarrow S$ that preserves
sums $0,+$ and multiplication, but not necessarily the unit. We define
a functor $\square \colon \op{\CRng}_{\klbullet} \rightarrow
\op{\PoSets}$ by $\square(R) = \Pred(R)$, the set of idempotents. For
a subunital map $f\colon R \rightarrow S$ with define $\square(f)
\colon \Pred(R) \rightarrow \Pred(S)$ by $\square(f)(e) = f(e) +
f(1)^{\bot}$. We see that $\square(f)(1) = 1$, so
Lemma~\ref{lem:grothendieck} applies.

\parpic[r][l]{%
\begin{minipage}{.5\columnwidth}
\begin{equation}
\label{diag:crng}
\vcenter{\xymatrix{
\int\square\ar[d]_{\dashv\;}^{\;\dashv}
   \ar@/_8ex/[d]^{\;\dashv}_{\begin{array}{c}\scriptstyle\mathrm{Quotient} \\[-.7em]
                        \scriptstyle(e\in R) \mapsto e^{\bot}R\end{array}} 
   \ar@/^8ex/[d]_{\dashv\;}^{\begin{array}{c}\scriptstyle\mathrm{Comprehension} \\[-.7em]
                        \scriptstyle(e\in R) \mapsto eR\end{array}} \\
\op{\CRng}_{\klbullet}\ar@/^4ex/[u]^(0.4){0\!}\ar@/_4ex/[u]_(0.4){\!1}
}}
\end{equation}
\end{minipage}}
Also in this case we have quotient and comprehension,
see~\eqref{diag:crng}.  Comprehension $\cmpr{R}{e}$ for an idempotent
$e\in R$ is given by the principal ideal $eR$, or equivalently the
ring of fractions $R[e^{-1}]$. The associated projection map $\pi_{e}
\colon R \rightarrow eR$ is given by $\pi_{e}(x) = ex$. For a
subunital map $f\colon R \rightarrow S$ with $1 \leq \square(f)(e) =
f(e) + f(1)^{\bot}$ we get $f(e) = f(1)$. The restriction
$\overline{f} \colon eR \rightarrow S$ of $f$ then satisfies
$\overline{f} \after \pi_{e} = f$, since $\overline{f}(\pi_{e}(x)) =
f(ex) = f(e)f(x) = f(1)f(x) = f(1x) = f(x)$.

We also show that quotients are given by $R/e = e^{\bot}R$, with
inclusion $\xi_{e} \colon e^{\bot}R \rightarrow R$ as subunital
quotient map.  Let $f\colon S \rightarrow R$ be a subunital map with
$e \leq \square(f)(0) = f(1)^{\bot}$. Hence $f(1) \leq e^{\bot}$ and
thus $e^{\bot}f(1) = f(1)$. We define $\overline{f} \colon S
\rightarrow e^{\bot}R$ as $\overline{f}(x) = f(x)$. Then:
$$\begin{array}{rcccccccccl}
\big(\xi_{e} \after \overline{f}\big)(x)
& = &
\xi_{e}(f(x)) 
& = &
e^{\bot} f(1\cdot x) 
& = &
e^{\bot} f(1) f(x) 
& = &
f(1) f(x) 
& = &
f(x).
\end{array}$$

\noindent Finally we notice that each idempotent $e\in R$ gives a
decomposition $R \cong eR \times e^{\bot}R = \cmpr{R}{e} \times
Q/e$. This decomposition is essential in the sheaf theory of
commutative rings, see~\cite[Chap.~IV]{Johnstone82}
or~\cite[Part~III]{Borceux94} for details. The instrument takes the
form $\instr_{e} \colon R\times R \rightarrow R$, and implicitly uses
this decomposition in: $\instr_{e}(x, y) = ex + e^{\bot}y$.

A similar example can be constructed for MV-modules, that is for
MV-algebras with a suitable $[0,1]$-scalar multiplication. They are
effect modules with a join $\vee$ (and then also meet $\wedge$)
interacting appropriately with the other structure. MV-modules are
also called Riesz MV-algebras, see~\cite{DiNolaL14}. The predicates on
an MV-module $A$ are the `sharp' elements $p\in A$ satisfying
$p^{\bot} \wedge p = 0$; they form a Boolean algebra. Comprehension
$\cmpr{A}{p}$ is $\downset p$ and quotient $A/p$ is $\downset
p^{\bot}$. Again, there is a decomposition $A \cong \downset p \times
\downset p^{\bot} = \cmpr{A}{p} \times A/p$, like for rings, see
also~\cite[6.4]{CignoliDM00}. Details will be elaborated elsewhere.

The opposite of the category of commutative $C^*$-algebra with
*-homomorphisms fits in this same pattern. We have already seen in
Section~\ref{sec:sets} that it has a quotient--comprehension chain,
because of the equivalence with the (extensive) category $\CH$ of
compact Hausdorff spaces.

\section{A Quantum Example}\label{sec:quant}
Von Neumann algebras
also yield
a quotient--comprehension chain, see~\eqref{diag:vn} below.
Strikingly, in this setting of quantum computation,
one quotient--comprehension chain gives us
the sequential product $a*b= \sqrt{a}b\sqrt{a}$
which is used to describe (sequential) measurement
on quantum systems~\cite{gudder2}.
%This not only gives us faith that the canonical
%description in the physics literature of measurement is the right
%one (at least, from a categorical perspective), 
%but it also enables us to define measurement abstractly (leading
%to the notion of telos, see Section~\ref{sec:telos}).
Since a rigorous treatment of the results
in this section
requires solid
understanding of functional analysis
we have 
collected the proofs and details
in a separate manuscript~\cite{WesterbaanW15}
and we permit ourselves  here an easygoing
narrative.

We model a quantum system by a \emph{von Neumann algebra}~$\mathscr{A}$
(see \cite{murray36,sakai71}).
A finite dimensional von Neumann algebra
is just a ring of matrices (closed under complex conjugation).
The reader is encouraged to keep this example in mind!
Roughly speaking
an element~$a$ of the von Neumann algebra~$\mathscr{A}$
(called an operator)
represents
both
an observable,
and the act of measuring it.
\emph{Qubits}
are modelled as~$2\times 2$ complex matrices over~$\C$.

Operators of the form~$a^*a$
are called \emph{positive}.
Their significance lies
in the fact that the linear maps $\varphi\colon \mathscr{A}\to \C$
with~$\varphi(1)=1$  which map positive operators to positive numbers
represent the \emph{states} of the system;
the number~$\varphi(a)$ for an operator~$a\in \mathscr{A}$
is the expectation value when measuring
observable~$a$ in state~$\varphi$.
We are only interested in states that are \emph{normal},
i.e.~preserve directed suprema of positive operators.
(Normality is only a concern
for infinite dimensional von Neumann algebras: 
a state on a ring of matrices is always normal.)
A computation which takes its input from a quantum system~$\mathscr{A}$
and ends up in~$\mathscr{B}$
is represented by a linear map $f\colon \mathscr{B}\to\mathscr{A}$
which is  \emph{positive} (maps positive operators 
to positive operators),
\emph{normal} (preserves directed suprema of positive operators)
and \emph{unital} ($f(1)=1$);
we say that
$f$ is a \emph{PNU-map}.
If the type of the map~$f$ surprises you, note that~$f$ allows us to 
transforms a normal state~$\varphi\colon \mathscr{A}\to \C$
on~$\mathscr{A}$
to a normal state~$\varphi\circ f$ on~$\mathscr{B}$.
The von Neumann algebra~$\C$ has only one state,
and so contains no data.
The state $\varphi\colon \mathscr{A}\to \C$
thus represents the computation without
input that initialises~$\mathscr{A}$
in state~$\varphi$.

The (parallel) composition
of two quantum systems
$\mathscr{A}$ and~$\mathscr{B}$
is represented by the tensor product~$\mathscr{A}\otimes\mathscr{B}$
of which the details are delicate.
One subtlety is that given computations (=PNU-maps) 
$f_1\colon \mathscr{A}_1\to\mathscr{B}_1$
and $f_2\colon \mathscr{A}_2\to\mathscr{B}_2$
their combination
$f_1\otimes f_2\colon\colon \mathscr{A}_1 \otimes \mathscr{A}_2
\to \mathscr{B}_1 \otimes \mathscr{B}_2$
need not be positive (i.e.~map positive operators to positive operators),
even when $f_2 \equiv \id_\mathscr{A} \colon \mathscr{A}\to \mathscr{A}$.

A PNU-map $f$ for which $f\otimes \id_\mathscr{A}$
is positive for every~$\mathscr{A}$
is called \emph{completely positive}~\cite{paulsen2002}.
Such maps, \emph{cPNU-maps} for short, 
are for our purposes the properly behaved quantum computations.
Let~$\vN$ denote the category
of cPNU-maps between von Neumann algebras.

One final detail:
as in the classical and probabilistic examples,
we need to consider partial maps to obtain a
quotient--comprehension chain. A partial quantum computation
from~$\mathscr{A}$ to~$\mathscr{B}$
is simply a completely positive normal 
linear map~$f\colon \mathscr{B}\to\mathscr{A}$
which is \emph{subunital}, i.e., $f(1)\leq 1$.
These maps between von Neumann algebras which we will call \emph{cPNsU-maps}
form a category~$\vNkl$.
Interestingly,
any (`partial') cPNsU-map $f\colon \mathscr{B}\to \mathscr{A}$
gives us a (`total') cPNU-map $g\colon \mathscr{B} \times \C \to \mathscr{A}$
via the equality~$g(b, \lambda)=f(b)+\lambda\cdot 1$.
This gives a bijection
between cPNsU-maps $\mathscr{B} \to \mathscr{A}$
and cPNU-maps $\mathscr{B} \times \C\to\mathscr{A}$.
In fact,
$\vNkl$ is isomorphic
to the Kleisli category
of the comonad~$(-)\times \C$ on the category~$\vN$.
Put differently,
$\vNopkl$
is isomorphic
to the Kleisli category
of the lift monad~$(-)+1$ on the opposite category~$\vNop$,
as is consistent with Definition~\ref{def:lift}.

The predicates on a quantum system
(represented by a von Neumann algebra~$\mathscr{A}$)
are the operators~$p$ in~$\mathscr{A}$
with~$0\leq p \leq 1$  called \emph{effects}.
The set of effects, $\Eff{\mathscr{A}}$,
is ordered by:~$p\leq q$ if $q-p$ is positive.
Note that~$1$ is the greatest and~$0$ is the least
element of~$\Eff{\mathscr{A}}$.
Given $p\in \Eff{\mathscr{A}}$
we write~$p^\perp :=1-p$.
The effects~$p$
for which~$p\wedge p^\perp = 0$
are called \emph{projections}.
It is notable that the projections
(in a von Neumann algebra) form
a complete lattice
while~$\Eff{\mathscr{A}}$
might not even be a lattice.
The least projection above an effect~$p$
is denoted by~$\ceil{p}$;
the greatest projection below~$p$ is denoted by~$\floor{p}$.

We can now get down to business.
Let~$\square\colon \vNopkl \to \op{\PoSets}$
be given by
$\square(\mathscr{A})=\Eff{\mathscr{A}}$
for every von Neumann algebra~$\mathscr{A}$
and~$\square (f)(p)=f(p^\perp)^\perp$
for every~$f\colon \mathscr{A}\to\mathscr{B}$
and~$p\in \mathscr{B}$.
The definition of~$\square$ is designed
to give us~$\square(f)(1)=1$
so that by Lemma~\ref{lem:grothendieck}
the forgetful functor $\int \square\to \vNopkl$
has a left adjoint~$0$ and right adjoint~$1$.
\begin{proposition}
\label{prop:quantum}
We have two more adjunctions
giving the chain~\eqref{diag:vn} below,
see~\cite{WesterbaanW15}.
\end{proposition}
% \begin{itemize}
% \item
\parpic[r][l]{%
\begin{minipage}{.7\columnwidth}
\begin{equation}
\label{diag:vn}
\vcenter{\xymatrix{
\int\square\ar[d]_{\dashv\;}^{\;\dashv}
   \ar@/_8ex/[d]^{\;\dashv}_{\begin{array}{c}\scriptstyle\mathrm{Quotient} \\[-.7em]
                   \scriptstyle (p\in\Eff{\mathscr{A}}) \mapsto 
   \ceil{p^\perp}\mathscr{A}\ceil{p^\perp}\end{array}} 
   \ar@/^8ex/[d]_{\dashv\;}^{\begin{array}{c}\scriptstyle\mathrm{Comprehension} \\[-.7em]
                   \scriptstyle(p\in\Eff{\mathscr{A}}) \mapsto 
   \floor{p}\mathscr{A}\floor{p}\end{array}} \\
   \vNopkl\ar@/^4ex/[u]^(0.4){0\!}\ar@/_4ex/[u]_(0.4){\!1}
}}
\end{equation}
\end{minipage}}
The comprehension functor
sends an effect $p\in\Eff{\mathscr{A}}$
to
the von Neumann algebra~$\floor{p}\mathscr{A}\floor{p}$
which has unit~$\floor{p}$.
The counit of the adjunction between~$1$ and comprehension
on~$p$
is the cPNsU-map $\pi_p\colon \mathscr{A}\to \floor{p}\mathscr{A}\floor{p}$
which sends~$a$ to~$\floor{p}a\floor{p}$.
% \item

The quotient functor
sends
an effect~$p$ of a von Neumann algebra~$\mathscr{A}$
to
the set of elements of~$\mathscr{A}$
of the form~$\ceil{p^\perp}a\ceil{p^\perp}$
which is denoted by~$\ceil{p^\perp}\mathscr{A}\ceil{p^\perp}$.
We should note
that~$\ceil{p^\perp}\mathscr{A}\ceil{p^\perp}$
is a
linear subspace
of~$\mathscr{A}$
which is
closed under multiplication, involution~$(-)^*$
and is closed
in the weak operator topology,
so that~$\ceil{p^\perp}\mathscr{A}\ceil{p^\perp}$
is itself (isomorphic to) a von Neumann algebra.
The unit of~$\ceil{p^\perp}\mathscr{A}\ceil{p^\perp}$
is~$\ceil{p^\perp}$
which might be different from the unit of~$\mathscr{A}$.
The unit of the adjunction between quotient and~$0$
on the effect~$p\in\mathscr{A}$
is the cPNsU-map~$\xi_p\colon 
\ceil{p^\perp}\mathscr{A}\ceil{p^\perp}\to \mathscr{A}$
which sends~$a$ to~$\smash{\sqrt{p^\perp} a \sqrt{p^\perp}}$.

% \end{itemize}
As in the probabilistic example,
we can form the following composites in~$\vNkl$.
\begin{equation}
\label{diag:dstinstr}
\xymatrix@R-2pc@C-.5pc{
	\ceil{p}\mathscr{A}\ceil{p}
& 
\mathscr{A}
\ar[l]_-{\pi_{\ceil{p}}} 
& 
\ceil{p}\mathscr{A}\ceil{p}
\ar[l]_-{\xi_{p^{\bot}}} 
&
\mathscr{A}
& 
\ceil{p}\mathscr{A}\ceil{p}
\ar[l]_-{\xi_{p^{\bot}}} 
& 
\mathscr{A}
\ar[l]_-{\pi_{\ceil{p}}} 
\\
\sqrt{p}a\sqrt{p} 
& 
\sqrt{p}a\sqrt{p}
\ar@{|->}[l] 
&
a
\ar@{|->}[l] 
&
\sqrt{p}a\sqrt{p} 
&
\ceil{p}a\ceil{p}
\ar@{|->}[l] 
  & 
  a
 \ar@{|->}[l]
}
\end{equation}
\noindent The map on the left is the identity if the predicate $p$ is
sharp. The map on the right is the `assert' map $\asrt_p$, which
yields, together with $\asrt_{p^{\bot}}$ the instrument $\instr_{p}
\colon \mathscr{A}\times \mathscr{A}\to\mathscr{A}$ in $\vNkl$ given
by 
\begin{equation*}
\instr_{p}(a,b) \ =\  \sqrt{p}\,a\,\sqrt{p}\,+\,\sqrt{1-p}\, b
\,\sqrt{1-p}
\end{equation*}
precisely as in~\cite{Jacobs14}. Hence we see how the
instrument map for measurement for von Neumann algebras is obtained
via the logical constructions of quotient and comprehension.%

%%%%%%%%%%%%%%%%%%%

\begin{myproof}[Proofsketch of Proposition~\ref{prop:quantum}]
\emph{(Comprehension)}\ 
We must show that given a von Neumann algebra~$\mathscr{A}$,
an effect~$p\in\mathscr{A}$,
and a map~$f\colon \mathscr{A}\to \mathscr{B}$
in~$\vNkl$ with~$f(p)=f(1)$
there is a unique map~$g\colon \floor{p}\mathscr{A}\floor{p}\to\mathscr{B}$
in~$\vNkl$
with~$g(\floor{p}b\floor{p})=f(b)$. 
Put~$g(b)=f(b)$; the difficulty it to show that~$f(\floor{p}b\floor{p})=f(b)$.
By a variant of Cauchy--Schwarz inequality for the completely positive
map~$f$ (see~\cite{paulsen2002}, exercise~3.4) 
\begin{equation*}
\|f(c^*d)\|^2 \ \leq\ \|f(c^*c)\|\cdot\|f(d^*d)\|
\qquad(c,d\in\mathscr{A})
\end{equation*}
we can reduce this problem to proving that~$f(\floor{p})=f(1)$,
that is, $f(\ceil{p^\perp})=0$.
Since~$\ceil{p^\perp}$
is the supremum of~$p^\perp\leq (p^\perp)^{\nicefrac{1}{2}}
\leq (p^\perp)^{\nicefrac{1}{4}}\leq\dotsb$
and~$f$ is normal,
$f(\ceil{p^\perp})$
is the supremum of the operators~$f(p^\perp)\leq f((p^\perp)^{\nicefrac{1}{2}})
\leq f((p^\perp)^{\nicefrac{1}{4}})\leq\dotsb$,
which all turn out to be zero by Cauchy--Schwarz  since~$f(p)=f(1)$. 
Thus~$f(\ceil{p^\perp})=0$,
and we are done.
Again,
for more details,
see~\cite{WesterbaanW15}.

\vspace{.2em}
\noindent
\emph{(Quotient)}\  We must show that
given a von Neumann algebra~$\mathscr{A}$, an effect~$p\in\mathscr{A}$,
and a map~$f\colon \mathscr{B} \to \mathscr{A}$
in~$\vNkl$ with~$f(1) \leq p^\perp$,
there is a unique~$g \colon \mathscr{B} 
            \to \ceil{p^\perp} \mathscr{A} \ceil{p^\perp}$
        in~$\vNkl$
such that~$\sqrt{p^\perp} g(b) \sqrt{p^\perp} = f(b)$.
If~$\sqrt{p^\perp}$ is invertible,
then we may define~$g(b) 
= \smash{(\sqrt{p^\perp})^{-1}\, f(b) \,(\sqrt{p^\perp})^{-1}}$,
and this works.
The proof is also straightforward if~$\sqrt{p^\perp}$ is pseudoinvertible
(=has norm-closed range).
The trouble is that in general~$\sqrt{p^\perp}$ is not (pseudo)invertible.
However, 
using the spectral theorem~\cite{halmos57}
we can find a sequence~$s_n$
(which converges ultraweakly to the (pseudo)inverse if it exists
and) for which~$g(b) = \uwlim_n s_n f(b) s_n$
exists and satisfies the requirements. 
For further details,
see~\cite{WesterbaanW15}. \myQED
\end{myproof}

\section{Conclusions}\label{sec:conc}

This paper uncovers a fundamental chain of adjunctions for quotient
and comprehension in many example categories of mathematical
structures, in particular von Neumann algebras. This in itself is
a discovery. Truly fascinating to us is the role that these adjunctions
play in the description of measurement instruments in these examples.
To our regret we are unable at this stage to offer a unifying
categorical formalisation, since in each of the examples there is
an equality connecting adjoints which are determined only
up-to-isomorphism. To be continued!

% \bibliographystyle{eptcs}
% \bibliography{common}

%%%% auxproof of the appendix
\auxproof{
%%%%

\appendix
\section{Details of the Quotient--Comprehension Chain for Extensive Categories}
In this section
we construct a quotient--comprehension chain
for all extensive categories with final object such as toposes.
\begin{definition}
A category~$\mathscr{E}$ is \keyword{extensive}
if it has all finite coproducts, 
a final object~$1$, and
pullbacks along coprojections of finite coproducts
such that in a diagram of the form
\begin{equation*}
\xymatrix{
	X\ar[d]^f \ar[r]^r &
	Z \ar[d]^h &
	Y \ar[l]_{s}\ar[d]^g \\
	A \ar[r]_{\kappa_1}& 
	A+B & 
	B \ar[l]^{\kappa_2}
}
\end{equation*}
both squares are pullbacks if and only if 
$\smash{\xymatrix{
	X\ar[r]^r &
	Z &
	Y \ar[l]_{s} 
}}$
is a coproduct diagram.
\end{definition}

\begin{remark}
We require that an extensive category
has a final object
contrary to~\cite{Carboni93}
so that we need not write ``extensive category with final object'' too often.
\end{remark}

\begin{definition}
\label{def:subobject}
Let~$\mathscr{E}$ be an extensive category.
Let~$X$ be an object of~$\mathscr{E}$.
\begin{enumerate}
\item
A \keyword{subobject} of~$X$
is a mono $m\colon S\to X$.
\end{enumerate}
Let~$m_1\colon S_1\to X$ and~$m_2\colon S_2\to X$
be subobjects of~$X$.
\begin{enumerate}[resume]
\item
\label{def:subobject-order}
We say that~$m_1$ is a \keyword{smaller than}~$m_2$, 
and write~$m_1\leq m_2$,
if there is an arrow~$j\colon S_1 \to S_2$
	such that $m_2\circ j = m_1$.
If~$j$ is an isomorphism,
we say that~$m_1$ and~$m_2$ are \keyword{equivalent},
and write~$m_1\approx m_2$.

\item
\label{def:subobject-complement}
If  $[m_1,m_2]\colon S_1+S_2\to X$ is an isomorphism
we say that~$m_1$ is \keyword{complemented} (by~$m_2$).

\item
\label{def:subobject-disjoint}
We say that~$m_1$ and~$m_2$
are \keyword{disjoint}
if the pullback of~$\smash{\xymatrix{S_1\ar[r]^{m_1} & X & S_2 \ar[l]_{m_2}}}$
	is~$0$.
\end{enumerate}
\end{definition}

It is easy to see that the relation~$\leq$
in subobjects
from Definition~\ref{def:subobject}\ref{def:subobject-order}
is reflexive and transitive,
and that the relation~$\approx$ on subobjects is an equivalence.

We shall now prove that any complemented subobject
has a unique complement. This requires some preparation.

\begin{lemma}
\label{lem:ec-coprojections-disjoint}
For any coproduct diagram 
$\smash{\xymatrix{A\ar[r]^{\kappa_1} & A+B & B\ar[l]_{\kappa_2}}}$
in an extensive category
the coprojections $\kappa_1$ and~$\kappa_2$ are disjoint and monic.
\end{lemma}
\begin{myproof}
See Proposition~2.6 of~\cite{Carboni93}.\myQED
\end{myproof}
\begin{corollary}
\label{cor:ec-complements-disjoint}
In an extensive category
subobjects $m_1$ and~$m_2$ are disjoint
if~$m_1$ is complemented
by~$m_2$.\myQED
\end{corollary}

\begin{lemma}
\label{lem:ec-complements}
Let~$\mathscr{E}$ be an extensive category.
Let~$X$ be an object of~$\mathscr{E}$.\\
Let~$m_1\colon S_1\to X$ and~$m_2\colon S_2 \to X$
be subobjects (of~$X$).
\begin{enumerate}
\item 
\label{lem:ec-complements-1}
If $m_1 \leq m_2$ and~$m_2 \leq m_1$,
then~$m_1 \approx m_2$.

\item
\label{lem:ec-complements-2}
If~$m_1$ and~$m_2$ complement the same subobject
(see Definition~\ref{def:subobject}\ref{def:subobject-complement})
then~$m_1\approx m_2$.
\end{enumerate}
\end{lemma}
\begin{myproof}
\ref{lem:ec-complements-1}\ 
Since~$m_1 \leq m_2$ 
there is $j_1\colon S_1\to S_2$
with $m_1 = m_2 \circ j_1$.
There is also an arrow~$j_2\colon S_2\to S_1$
with~$m_2 = m_1 \circ j_2$ as~$m_2 \leq m_1$.
Then~$m_1 = m_1 \circ j_2 \circ j_1$.
As~$m_1$ is mono,
this entails~$j_2 \circ j_1 = \id$.
By a similar reasoning, $j_1 \circ j_2 = \id$.
Thus~$j_1$ is an isomorphism,
and so~$m_1 \approx m_2$.

\ref{lem:ec-complements-2}\ 
Let~$m\colon S\to X$
be a subobject
such that~$m$ is complemented by~$m_1$
and~$m$ is complemented by~$m_2$.
Without loss of generality,
we may write~$X=S+S_1$
and~$m_1=\kappa_2 \colon S_1 \to S+S_1$
and~$m=\kappa_1\colon S\to S+S_1$.
Consider
the following diagram.
\begin{equation*}
\xymatrix{
0
\ar[r]
\ar[d]
&
S_2
\ar[d]_{m_2}
&
T
\ar[d]^j
\ar[l]_{\varphi}
\\
S
\ar[r]_{\kappa_1}
&
S+S_1
&
S_1
\ar[l]^{\kappa_2}
}
\end{equation*}
Let us take the right square to be the pullback of~$m_2$ and~$\kappa_2$
(which exists, since~$\kappa_2$ is a coprojection).
Since~$m\equiv\kappa_1$ and~$m_2$ are disjoint
(by Corollary~\ref{cor:ec-complements-disjoint})
the left square is a pullback as well.
Thus, 
since~$\mathscr{E}$ is extensive,
it follows that
$\smash{\xymatrix{0\ar[r]&S_1 & T\ar[l]_\varphi}}$
is a coproduct diagram.
But then~$\varphi$ is an isomorphism.
Thus we see that~$m_2 \leq m_1\equiv \kappa_2$ 
via $j\circ\varphi^{-1}\colon S_2\to S_1$. 

By a similar reasoning, we get~$m_1 \leq m_2$.
Thus~$m_1\approx m_2$ by part~\ref{lem:ec-complements-1}.\myQED
\end{myproof}
\begin{notation}
Let~$m\colon S\to X$
be a complemented subobject in an extensive category~$\mathscr{E}$.

By Lemma~\ref{lem:ec-complements}\ref{lem:ec-complements-2},
$m$~is complemented by a unique subobject (up to equivalence).
We will denote this subobject by~$m^\perp\colon S^\perp\to X$
and call it the \keyword{complement} of~$m$.
\end{notation}

We now turn to the definition
of the functor $\square\colon \mathscr{E}_\klbullet\to \op{\PoSets}$
(where~$\mathscr{E}_\klbullet$ is the Kleisli
category of the lift monad $(-)+1$)
which will give us our quotient--comprehension chain.
We can already unveil
that~$\square(X)$ will be the poset of complemented subobjects of~$X$
modulo~$\approx$,
but the definition of the action of~$\square$ an arrows
requires us to dig up some more facts about extensive categories first.
\begin{lemma}
\label{lem:ec-pullbacks}
Let $\mathscr{E}$ be an extensive category.
Consider a triple of diagrams of the following shape.
\begin{equation*}
\xymatrix{
P_1
\ar[r]^{q_1}
\ar[d]_{p_1}
&
B_1
\ar[d]^{b_1}
\\
A_1
\ar[r]_{a_1}
&
C_1
}
\qquad
\xymatrix{
P_2
\ar[r]^{q_2}
\ar[d]_{p_2}
&
B_2
\ar[d]^{b_2}
\\
A_2
\ar[r]_{a_2}
&
C_2
}
\qquad
\xymatrix{
P_1+P_2
\ar[r]^{q_1+q_2}
\ar[d]_{p_1+p_2}
&
B_1+B_2
\ar[d]^{b_1+b_2}
\\
A_1+A_2
\ar[r]_{a_1+a_2}
&
C_1+C_2
}
\end{equation*}
The two diagrams on the left  are pullbacks
if and only if  the diagram on the right is a pullback.
\end{lemma}
\begin{myproof}
$(\Longrightarrow)$\ \emph{(Existence)}\ 
Let~$\smash{\xymatrix{A_1+A_2 & D \ar[r]|{v}\ar[l]|{u} & B_1+B_2}}$
with~$(a_1+a_2)\circ u = (b_1+b_2)\circ v$ be given.
We must show that there is a unique~$w \colon D\to P_1+P_2$
such that $(p_1+p_2)\circ w = u$
and $(q_1+q_2)\circ w = v$.

Write~$\xi :=(a_1+a_2)\circ u = (b_1+b_2)\circ v$
and consider the diagram
\begin{equation*}
\xymatrix{
	D_1\ar[d]^{\xi_1} \ar[r] &
	D \ar[d]^\xi &
	D_2 \ar[l]\ar[d]^{\xi_2} \\
	C_1 \ar[r]_{\kappa_1}& 
	C_1+C_2 & 
	C_2 \ar[l]^{\kappa_2}
}
\end{equation*}
where~$D_1$, $D_2$, $\xi_1$, $\xi_2$, \ldots~are
chosen in such a way that both squares are pullbacks.
Then since~$\mathscr{E}$ is extensive
$\smash{\xymatrix{ D_1 \ar[r] & D & D_2\ar[l] } }$
is a coproduct diagram.
So without loss of generality,
we assume that $D\equiv D_1 + D_2$
and~$\xi=[\xi_1,\xi_2]$.
Now consider this diagram:
\begin{equation*}
\xymatrix{
	D_1\ar@/_1em/[dd]_{\xi_1} \ar[r]^{\kappa_1} &
	D_1+D_2 \ar[d]^v &
	D_2 \ar[l]_{\kappa_2}\ar@/^1em/[dd]^{\xi_2} \\
	B_1 \ar[r]^{\kappa_1}\ar[d]^{b_1} &
	B_1+B_2 \ar[d]^{b_1+b_2} &
	B_2 \ar[l]_{\kappa_2}\ar[d]_{b_2}\\
	C_1 \ar[r]_{\kappa_1}& 
	C_1+C_2 & 
	C_2 \ar[l]^{\kappa_2}
}
\end{equation*}
The square $B_1(B_1+B_2)(C_1+C_2)C_1$
is a pullback
so
there is a unique $v_1\colon D_1 \to B_1$
such that $b_1\circ v_1 = \xi_1$
and~$\kappa_1\circ v_1 = v\circ \kappa_1$.
Doing the same on the other side,
we obtain~$v_2\colon D_2 \to B_2$
with $b_2 \circ v_2 = \xi_2$
and~$\kappa_2 \circ v_2 = v\circ \kappa_2$.
It follows that~$v=v_1 + v_2$.
Similarly, we get $u\equiv u_1+u_2$ where  $u_1\colon D_1\to A_1$
and~$u_2\colon D_2\to A_2$.

Since~$(a_1+a_2)\circ u = (b_1+b_2)\circ v$
we get
$a_1\circ u_1 = b_1 \circ v_1$.
Since $\smash{\xymatrix{A_1 & P_1 \ar[r]|{q_1} \ar[l]|{p_1} & B_1}}$
is the pullback of
$\smash{\xymatrix{A_1 \ar[r]|{a_1} & P_1 & B_1 \ar[l]|{b_1} }}$
there is a (unique) $w_1\colon D_1 \to P_1$
such that $p_1\circ w_1 = u_1$ and $q_1\circ w_1 = v_1$.

Similarly, we get a unique arrow $w_2\colon D_2\to P_2$
such that $p_2\circ w_2 = u_2$ and $q_2\circ w_2 = v_2$.
Then $(p_1+p_2)\circ (w_1+w_2) = u_1+u_2 \equiv u$
and $(q_1+q_2)\circ (w_1+w_2) = v_1+v_2 \equiv v$.

\emph{(Uniqueness)}\ 
Suppose that an arrow~$w\colon D_1+D_2\to P_1+P_2$
with $(p_1+p_2)\circ w = u_1+u_2$
and $(q_1+q_2)\circ w = v_1+v_2$ is given.
We must show that~$w=w_1+w_2$.

Consider the following diagram.
\begin{equation*}
\xymatrix{
	D_1\ar@/_1em/[dd]_{u_1} \ar[r]^{\kappa_1} &
	D_1+D_2 \ar[d]^w &
	D_2 \ar[l]_{\kappa_2}\ar@/^1em/[dd]^{u_2} \\
	P_1 \ar[r]^{\kappa_1}\ar[d]^{p_1} &
	P_1+P_2 \ar[d]^{p_1+p_2} &
	P_2 \ar[l]_{\kappa_2}\ar[d]_{p_2}\\
	A_1 \ar[r]_{\kappa_1}& 
	A_1+A_2 & 
	A_2 \ar[l]^{\kappa_2}
}
\end{equation*}
Since the square in the bottom left corner is a pullback
we get $w_1'\colon D_1\to P_1$
with $\kappa_1 \circ w_1' = w\circ \kappa_1$
and~$p_1\circ w_1'= u_1$.
Also, we get~$w_2'\colon D_2\to P_2$
with $\kappa_2 \circ w_2' = w\circ \kappa_2$
and~$p_2\circ w_2'= u_2$.
Then~$w_1'+w_2'=w$.

Since~$(p_1+p_2)\circ w = u_1+u_2$
we get~$p_1 \circ w_1' = u_1$
and since $(q_1+q_2)\circ w = v_1+v_2$
we get~$q_1\circ w_1' = v_1$.
It follows that~$w_1=w_1'$.
Similarly, $w_2=w_2'$.
Hence~$w=w_1+w_2$.

$(\Longleftarrow)$\ 
We leave this to the reader.\myQED
\end{myproof}

\begin{corollary}
\label{cor:ec-e}
In an extensive category
a diagram of the following shape is a pullback diagram.
\begin{equation*}
\xymatrix{
X+A
\ar[r]^{f+\id}
\ar[d]_{\id+g}
&
Y+A
\ar[d]^{\id+g}
\\
X+B
\ar[r]_{f+\id}
&
Y+B
}
\end{equation*}
\end{corollary}

\begin{corollary}
\label{cor:ec-sum-pullbacks}
Given an object~$A$ of an extensive category~$\mathscr{E}$,
$(-)+A\colon\mathscr{E}\to\mathscr{E}$
preserves pullbacks.
\end{corollary}
\begin{corollary}
\label{cor:ec-sum-monos}
If~$m_1$ and~$m_2$
are monos in an extensive category,
then~$m_1+m_2$ is mono too.
\end{corollary}
\begin{myproof}
Use Lemma~\ref{lem:ec-pullbacks}
and the fact that
$f\colon X\to Y$ is mono
iff the following is a pullback.
\begin{equation*}
\xymatrix{
X\ar[r]^\id \ar[d]_\id & X\ar[d]^f \\ X\ar[r]_f & Y
}
\end{equation*}
\end{myproof}
\begin{corollary}
\label{cor:ec-complemented-subobject-sum}
If~$m_1$ and~$m_2$
are complemented subobjects in an extensive category,
then $m_1+m_2$ is a complemented subobject,
and~$(m_1+m_2)^\perp = m_1^\perp + m_2^\perp$.
\end{corollary}
\begin{myproof}
Use Corollary~\ref{cor:ec-sum-monos}
to see that~$m_1+m_2$ is a subobject.
We leave it to the reader to prove that~$m_1+m_2$ is complemented by
$m_1^\perp + m_2^\perp$.\myQED
\end{myproof}

\begin{definition}
Let~$f\colon X\to Y$ be an arrow in an extensive category~$\mathscr{E}$.
Let~$m\colon S\to Y$ be a complemented subobject.
Consider the following pullbacks
--- which exist since~$m$ and~$m^\perp$
are coprojections.
\begin{equation*}
\xymatrix{
	f^*(S)\ar[d] \ar[r]^{f^*(m)} &
	X \ar[d]^f &
	f^*(S^\perp) \ar[l]_{f^*(m^\perp)} \ar[d] \\
	S \ar[r]_{m}& 
	Y & 
	S^\perp \ar[l]^{m^\perp}
}
\end{equation*}
Since~$\mathscr{E}$ is extensive
we see that
$\xymatrix{
	f^*(S)\ar[r]^{f^*(m)} &
	X  & f^*(S^\perp) \ar[l]_{f^*(m^\perp)}}$
is a coproduct diagram,
and thus~$f^*(m)$ 
is a complemented subobject of~$X$ with $f^*(m)^\perp = f^*(m^\perp)$.
We call~$f^*(m)$ the \keyword{pullback of~$m$ along~$f$}.
\end{definition}

\begin{lemma}
\label{lem:ec-arrows-into-zero}
In an extensive category
any morphism~$X\to 0$
is an isomorphism.
\end{lemma}
\begin{myproof}
See Proposition~2.8 of~\cite{Carboni93}.\myQED
\end{myproof}

\begin{corollary}
In an extensive category
\begin{enumerate}
\item the unique arrow from~$0$ into an object~$X$ is a mono,
and complemented by~$\id\colon X\to X$;
\item
the pullback of $0\rightarrow Y$
along an arrow $f\colon X\rightarrow Y$ is~$0\rightarrow X$.\myQED
\end{enumerate}
\end{corollary}

\begin{definition}
Let~$\mathscr{E}$ be an extensive category.
\begin{enumerate}
\item
Let~$\mathscr{E}_\klbullet$
be the Kleisli category of the lift monad~$(-)+1$ on~$\mathscr{E}$.

\item
For an object~$X$ of~$\mathscr{E}$
let~$\square(X)$
be the poset of complemented subobjects on~$X$ modulo~$\approx$.

\item
Let $f\colon X\to Y+1$ be an arrow in~$\mathscr{E}$.
We define a map~$\square(f)\colon \square(Y)\to \square(X)$.

Let~$m\colon S\to Y$ be a complemented subobject.
Then~$m+1$
is a complemented subobject
by Corollary~\ref{cor:ec-complemented-subobject-sum}.
Let~$\square(f)(m) := f^*(m+1)$.
That is,
it is given by the following pullback.
\begin{equation*}
\xymatrix{
	\bullet
\ar[rr]^{\square(f)(m)}
\ar[d]
&&
X \ar[d]^f
\\
S+1\ar[rr]_{m+1} 
&&Y+1
}
\end{equation*}
\end{enumerate}
\end{definition}

\begin{proposition}
\label{prop:ec-square}
Let~$\mathscr{E}$ be an extensive category.
Let~$X$ be an object of~$\mathscr{X}$.
\begin{enumerate}
\item
\label{prop:ec-square-1}
$0\rightarrow X$
is the smallest element of~$\square(X)$, and
$\id\colon X\to X$
is the greatest element of~$\square(X)$.
\item
\label{prop:ec-square-2}
Let $f\colon X\to Y+1$ be from~$\mathscr{E}$.
Then~$\square(f)\colon \square(Y)\to\square(X)$
is order preserving, and $\square(f)(\id)=\id$.

\item 
\label{prop:ec-square-3}
We get a functor~$\square\colon \mathscr{E}_{\klbullet}\to\op{\PoSets}$. 
\end{enumerate}
\end{proposition}
\begin{myproof}
We leave~\ref{prop:ec-square-1} and~\ref{prop:ec-square-2}
to the reader.

\ref{prop:ec-square-3}\ 
\emph{(Identity)}\ 
Let~$X$ be an object of~$\mathscr{E}$.
Then~$\kappa_1\colon X\to X+1$ is the identity in~$\mathscr{E}_\klbullet$
on~$X$.
Let~$m\colon S\to X$ be a complemented subobject.
We must show that~$\square(\kappa_1)(m)=m$.
And indeed it is since the following diagram is a pullback
because~$\mathscr{E}$ is an extensive category.
\begin{equation*}
\xymatrix{
S \ar[r]^m  \ar[d]_{\kappa_1}& X \ar[d]^{\kappa_1} \\
S+1 \ar[r]_{m+1}  & X+1
}
\end{equation*}

\emph{(Composition)}\ 
Let~$f\colon X\to Y+1$
and~$g\colon Y\to Z+1$ from~$\mathscr{E}$ be given.
Let~$m\colon W\to Z$ be a complemented subobject.
Write~$\ell := \square(g)(m)$ and~$k := \square(f)(\ell)$.
We must show that~$\ell = \square(g\klafter f)$,
that is, we must show that the outer rectangle in the 
diagram below is a pullback.
\begin{equation*}
\xymatrix{
U
\ar[r]
\ar[d]_{k}
&
V+1
\ar[r]
\ar[d]_{\ell+1}
&
W+1+1
\ar[r]^{\id+[1,1]}
\ar[d]_{m+1+1}
&
W+1
\ar[d]^{m+1}
\\
X
\ar[r]^f
\ar@/_1em/[rrr]_{g\klafter f}
&
Y+1
\ar[r]^{g+1}
&
Z+1+1
\ar[r]^{\id+[1,1]}
&
Z+1
}
\end{equation*}
The first square is a pullback by definition of~$k=\square(f)(\ell)$.
The second square is a pullback by definition of~$\ell=\square(f)(\ell)$
and Corollary~\ref{cor:ec-sum-pullbacks}.
The third square is a pullback by Corollary~\ref{cor:ec-e}.

Hence the outer rectangle is a pullback as well.\myQED
\end{myproof}

\begin{proposition}
\label{prop:ec-comprehension}
Let~$m\colon S\to X$ be a complemented
subobject of an extensive category~$\mathscr{E}$.

\parpic[r][l]{%
\begin{minipage}{.2\columnwidth}
\begin{equation*}
\xymatrix{
X
&
S\ar[l]|\circ_{\varepsilon_m}
\\
&
Y\ar[lu]|\circ^{f} \ar[u]|\circ_{\overline{f}}
}
\end{equation*}
\end{minipage}}
\noindent
Then~$\varepsilon_m\colon S\to X+1$ given by~$\varepsilon_m = \kappa_1\circ m$
has the following universal property.
We have~$\square(\varepsilon_m)(m) = \id_S$,
and for every arrow $f\colon Y\to X+1$
with~$\square(f)(m) \approx \id_Y$
there is a unique arrow~$\overline{f}\colon Y\to S+1$
with~$\varepsilon_m\klafter\overline{f}=f$.
\end{proposition}
\begin{myproof}
To prove that~$\square(\varepsilon_m)(m)=\id_S$,
we must show that the outer rectangle below
below is a pullback.
\begin{equation*}
\xymatrix{
S \ar[d]_\id \ar[r]^\id&
S \ar[d]_m \ar[r]^{\kappa_1}&
S+1 \ar[d]^{m+1} 
\\
S\ar[r]_m&
X\ar[r]_{\kappa_1}&
X+1
}
\end{equation*}
It suffices to show that both squares are pullbacks.
The first square is a pullback because~$m$ is mono.
The second square is a pullback since~$\mathscr{E}$ is extensive.
Hence~$\square(\varepsilon_m)(m)=\id_S$.

Let $f\colon Y\to X+1$
with~$\square(f)(m) \approx \id_Y$
be given.
We must show there is a unique arrow~$g\colon Y\to S+1$
with~$\varepsilon_m\klafter g=f$.

\emph{(Uniqueness)}\ 
Let~$g_1,g_2\colon Y\to S+1$
with~$\varepsilon_m \klafter g_1 =\varepsilon_m \klafter g_2 = f$
be given.
We must show that~$g_1=g_2$.
Note that $\varepsilon_m \klafter g_1 = (m+1)\circ g_1$,
and $\varepsilon_m \klafter g_1 = (m+1)\circ g_2$.
Thus $(m+1)\circ g_1 = (m+1)\circ g_2$.
This entails~$g_1=g_2$
since~$m+1$ is mono (by Corollary~\ref{cor:ec-sum-monos}).

\emph{(Existence)}\ 
Since~$\square(f)(m)\approx \id_Y$
we have the following pullback square.
\begin{equation*}
\xymatrix{
Y \ar[r]^g
\ar[d]_{\id}
&
S+1
\ar[d]^{m+1}
\\
Y
\ar[r]_f
&
X+1
}
\end{equation*}
Then $\varepsilon_m\klafter g = (m+1)\circ g = f$,
so~$g$ is the sought arrow.\myQED

\end{myproof}

\begin{proposition}
\label{prop:ec-quotient}
Let~$m\colon S\to X$ be a
subobject of an extensive category~$\mathscr{E}$
complemented by~$m^\perp\colon S^\perp \to X$.

\parpic[r][l]{%
\begin{minipage}{.2\columnwidth}
\begin{equation*}
\xymatrix{
X
\ar[r]|\circ^{\eta_m}
\ar[rd]|\circ_{f} 
&
S^{\perp}
\ar[d]|\circ^{\overline{f}}
\\
&
Y
}
\end{equation*}
\end{minipage}}
\noindent
Then~$\eta_m\colon X\to S^\perp+1$ given 
by~$\eta_m = (\id+!)\circ [m^\perp,m]^{-1}$
has the following universal property.
We have~$m=\square(\eta_m)(?_{S^\perp})$
where $?_S\colon 0\to {S^\perp}$,
and for every arrow $f\colon Y\to X+1$
with~$m \leq \square(f)(?_Y)$
there is a unique arrow~$\overline{f}\colon S^\perp \to Y+1$
with~$\overline{f}\klafter \eta_m=f$.
\end{proposition}
\begin{myproof}
To prove that~$m=\square(\eta_m)(?)$
we must show that the outer rectangle below is a pullback.
\begin{equation*}
\xymatrix{
S 
\ar[d]_m
\ar[r]^\id
&
S
\ar[d]_{\kappa_2}
\ar[r]^{\kappa_2\circ !}
&
0+1
\ar[d]^{?+1}
\\
X
\ar[r]_{[m^\perp,m]^{-1}}
&
S^\perp + S
\ar[r]_{\id+!}
&
S^\perp+1
}
\end{equation*}
It suffices to show that both squares are pullbacks.
The left square is a pullback 
since~$\smash{\xymatrix{S \ar[r]^m& X& S^\perp \ar[l]_{m^\perp}}}$
is a coproduct diagram.
To see that the right square is a pullback is even easier.

Let $f\colon Y\to X+1$
with~$m \leq \square(f)(?_Y)$
be given.
We must show
there is a unique arrow~$\overline{f}\colon S^\perp \to Y+1$
with~$\overline{f}\klafter \eta_m=f$.

\emph{(Existence)}\ 
Write~$n:= \square(f)(m)\colon Q\to X$
Since~$m\leq \square(f)(m)$
there is $j\colon S\to Q$
with~$m= n\circ j$.
It is easy to 
see that~$j$ is the pullback of~$m$ along~$n$
using the fact that~$m=n\circ j$ and~$n$ is mono.
Thus~$j$ is a complemented subobject of~$Q$.
So we have the following pullbacks.
\begin{equation*}
\xymatrix{
R
\ar[r]^{j^\perp}
\ar[d]
&
Q
\ar[d]^{n}
&
S
\ar[l]_{j}\ar[d]^{\id}
\\
S^\perp
\ar[r]_{m^\perp}
&
X
&
S
\ar[l]^{m}
}
\end{equation*}
To simply the situation,
we assume (without loss of generality)
that~$X\equiv Q^\perp+R+S$,
and also that
$Q\equiv R+S$ and~$Q^\perp+R\equiv S^\perp$
and that
$m\equiv\kappa_3\colon S\to Q^\perp+R+S$
and~$n\equiv \kappa_2\colon R+S\to Q^\perp+(R+S)$
and~$\eta_m\equiv\id+\id+!\colon Q^\perp + R + S \to Q^\perp + R + 1$.

We take the pullback
of~$\smash{\xymatrix{Y\ar[r]^{\kappa_1}& Y+1& 1\ar[l]_{\kappa_2}}}$
along~$f$ and obtain the following diagram,
where we have used the fact that~$\square(f)(?)=n\equiv \kappa_2\colon R+S
\to Q^\perp+(R+S)$.
\begin{equation*}
\xymatrix{
Q^\perp
\ar[r]^-{n^\perp}
\ar[d]_{t}
&
Q^\perp+R+S
\ar[d]|{f\equiv t+!}
&
R+S
\ar[l]_-{n}
\ar[d]^{!}
\\
Y
\ar[r]_-{\kappa_1}
&
Y+1
&
1
\ar[l]^-{\kappa_2}
}
\end{equation*}
Define~$g:= t+!\colon Q^\perp+R\to Y+1$.
Then a quick calculation yields $g \klafter \eta_m = t+!=f$.

\emph{(Uniqueness)}\ 
Let~$g\colon Q^\perp+R\to Y+1$
such that~$f=g\klafter \eta_m$ be given.
We must show that~$g=t+!$.
From~$f=g\klafter \eta_m$
one easily deduces the following diagram commutes.
\begin{equation*}
\xymatrix{
(Q^\perp+R)+S 
\ar[r]^-{g+!}
\ar[d]_{\id}
&
Y+1+1
\ar[d]^{\id+[1,1]}
\\
Q^\perp + (R+S)
\ar[r]_-{t+!}
&
Y+1
}
\end{equation*}
If we precompose with~$\kappa_1\colon Q^\perp+R \to Q^\perp+R+S$
we get~$g=t+!$, as required. \myQED

\end{myproof}

We summarise Proposition~\ref{prop:ec-comprehension} and 
Proposition~\ref{prop:ec-quotient}
by the following quotient--comprehension chain.
\begin{theorem}
There is a chain of adjunctions
as shown to the right below.

\parpic[r][l]{%
\begin{minipage}{.6\columnwidth}
\begin{equation}
\vcenter{\xymatrix{
\int\square\ar[d]_{\dashv\;}^{\;\dashv}
   \ar@/_8ex/[d]^{\;\dashv}_{\begin{array}{c}\scriptstyle\mathrm{Quotient} \\[-.7em]
                        \scriptstyle (m\colon S\to X) \mapsto S^\perp \end{array}} 
   \ar@/^8ex/[d]_{\dashv\;}^{\begin{array}{c}\scriptstyle\mathrm{Comprehension} \\[-.7em]
                        \scriptstyle(m\colon S\to X) \mapsto S\end{array}} \\
\mathscr{E}_{\klbullet}\ar@/^4ex/[u]^(0.4){0\!}\ar@/_4ex/[u]_(0.4){\!1}
}}
\end{equation}
\end{minipage}}
\noindent
Let~$m\colon S\to X$ be a complemented subobject
(and thus an object in~$\int\square$).

The counit~$\varepsilon_m \colon S\longrightarrow X+1$
of the adjunction between~$1$ and comprehension
at~$m$
is given by~$\varepsilon_m=\kappa_1\circ m$.\\
The unit~$\eta_m\colon X\to S^\perp+1$
of the adjunction between~$0$ and quotient is
given by~$\eta_m = (\id+!)\circ [m^\perp,m]^{-1}$.
\end{theorem}
\begin{myproof}
We leave this to the reader.\myQED
\end{myproof}

\section{$\op{\CRng}$ is an Extensive Category}

In this section,
we will show that the opposite of the category~$\op{\CRng}$
of commutative rings with homomorphisms
is extensive
(see Proposition~\ref{prop:cr-extensive}).
Recall that the product of $A$ and~$B$ in~$\CRng$
is the Cartesian product~$A\times B$
with the coordinatewise operations.

\begin{lemma}
\label{lem:cr-pushout}
In~$\CRng$
any homomorphism of the form $f\colon A \times B\rightarrow C$ has a pushout 
along~$\pi_1\colon A \times B\to A$.
\end{lemma}
\begin{myproof}
Consider the set~$I:= \{\ c_1 \,f(0,b_1)+ \dotsb + c_n\, f(0,b_n)\colon\ 
c_1,\dotsc,c_n\in C,\ b_1,\dotsc,b_n\in B\ \}$
of~$C$;
it is the ideal generated by $\{\,f(0,b)\colon\, b\in B\,\}$.

\parpic[r][r]{\begin{minipage}{.2\columnwidth}
\begin{equation*}
\xymatrix{
A\times B
\ar[r]^{\pi_1}
\ar[d]_f
&
A
\ar[d]^h
\\
C
\ar[r]_q
&
C/I
}
\end{equation*}
\end{minipage}}
\noindent
Note that the map~$\pi_1\colon A\times B\to A$
is surjective
and has kernel~$\{\,(0,b)\colon b\in B\,\}$.
Since~$f(0,b)\in I$ for all~$b\in B$
there is a unique homomorphism $h\colon A\to C/I$ such that
the diagram on the right commutes
where~$q\colon C\to C/I$ is the quotient map.
We claim that this is a pushout diagram
(and so~$q$ is the pushout of~$\pi_1$ along~$h$).

Let~$\smash{\xymatrix{C \ar[r]^\gamma& D & A\ar[l]_\alpha}}$
with~$\alpha\circ \pi_1 = \gamma\circ f$ be given.
We must show that there is a unique homomorphism
$\xi\colon C/I \to D$ such that~$\xi\circ h = \alpha$
and~$\xi\circ q = \gamma$.

\emph{(Existence)}\ 
We claim that~$\gamma(x)=0$ for all~$x\in I$.
Let~$b\in B$ be given.
Since~$\gamma$ is a homomorphism
it suffices to show that~$f(0,b)=0$.
We have $\gamma(f(0,b))= \alpha(\pi_1(0,b))=\alpha(0)=0$.

Since~$\gamma(x)=0$ for all~$x\in I$,
there is a unique homomorphism $\overline{\gamma}\colon C/I\to D$
with~$\overline{\gamma}\circ q = \gamma$. 
It remains to be checked that~$\overline{\gamma}\circ h = \alpha$.
Let~$a\in A$ be given.
Then~$\alpha(a)=\alpha(\pi_1(a,0)) = \gamma(f(a,0)) 
= \overline{\gamma}(q(f(a,0)))
= \overline{\gamma}(h(\pi_1(a,0))) 
= \overline{\gamma}(h(a))$.

\emph{(Uniqueness)}\ 
follows easily from the fact that~$q$ is surjective.\myQED
\end{myproof}

\begin{corollary}
\label{cor:cr-product-pushout-1}
In~$\CRng$,
the pushout of~$\pi_1\colon A\times B\to A$
along~$f\times g\colon A\times B\to X\times Y$
is~$\pi_1\colon X\times Y\to X$.
\end{corollary}
\begin{myproof}
Let~$I$ be the ideal generated by~$\{\,(f\times g)(0,b)\colon\, b\in B\,\}$.
By the proof of Lemma~\ref{lem:cr-pushout}
it suffices to show 
that~$I$ is the kernel of~$\pi_1$.
Note that~$I$ contains~$(0,1) \equiv (0,g(1))$.
Thus~$I=\{\,(0,y)\colon\,y\in Y\,\}$,
which is exactly the kernel of~$\pi_1\colon X\times Y\to X$.\myQED
\end{myproof}

\begin{lemma}
\label{lem:cr-ideals}
Let~$f\colon A\times B\to C$ be an arrow in~$\CRng$. Then
$I+J=C$ and $I\cap J =\{0\}$
where
\begin{alignat*}{3}
I\ &:=\  \{\ c_1 \,f(0,b_1)\,+\, \dotsb \,+\, c_n\, f(0,b_n)\colon\ 
c_1,\dotsc,c_n\in C,\ b_1,\dotsc,b_n\in B\ \} \\
J\ &:=\  \{\ d_1 \,f(a_1,0)\,+\, \dotsb \,+\, d_n\, f(a_m,0)\colon\ 
d_1,\dotsc,d_m\in C,\ a_1,\dotsc,a_m\in A\ \}.
\end{alignat*}
\end{lemma}
\begin{myproof}
Let~$c\in C$.
Then~$c=c\cdot 1 = c\cdot f(1,1)
= c\cdot f(0,1) =+ c\cdot f(1,0)$.
Since~$c\cdot f(0,1)\in I$ and~$c\cdot f(1,0)\in J$
we see that~$c\in I+J$.  Thus~$C=I+J$.

Let~$x\in I\cap J$ be given.
We must show that~$x=0$.
Since~$x = x\cdot f(0,1) + x\cdot f(1,0)$
it suffices to show that~$x\cdot f(0,1)=0$ and~$x\cdot f(1,0)=0$.
Since~$x\in I\cap J\subseteq J$
we can write $x\equiv \sum_j d_j f(a_j,0)$ for~$d_j\in D$ and~$a_j\in A$.
Then~$x\cdot f(0,1) = \sum_j d_j f(\,(a_j,0)\,\cdot\, (0,1)\,)
= 0$.
Similarly, $x\cdot f(1,0) = 0$. Thus $x=0$.\myQED
\end{myproof}

\begin{corollary}
\label{cor:cr-product-pushout-2}
If in a diagram in~$\CRng$
of the form
\begin{equation}
\label{eq:cr-extensive}
\xymatrix{
A 
\ar[d]_{s}
&
A\times B
\ar[l]_{\pi_1}
\ar[r]^{\pi_2}
\ar[d]_f
&
B
\ar[d]^t
\\
X
&
C
\ar[l]^{q}
\ar[r]_r
&
Y
}
\end{equation}
both squares are pushouts,
then
 $\smash{\xymatrix{X & C\ar[l]_{q} \ar[r]^r & Y}}$
is a product diagram.
\end{corollary}
\begin{myproof}
By Lemma~\ref{lem:cr-pushout} we can assume without loss of generality
that Diagram~\eqref{eq:cr-extensive}
has the form
\begin{equation*}
\xymatrix{
A 
\ar[d]_{s}
&
A\times B
\ar[l]_{\pi_1}
\ar[r]^{\pi_2}
\ar[d]_f
&
B
\ar[d]^t
\\
C/I
&
C
\ar[l]^{q}
\ar[r]_r
&
C/J
}
\end{equation*}
where~$I$ and~$J$ are as defined in Lemma~\ref{lem:cr-ideals}
and~$q$ and~$r$ are quotient maps.
Since~$I+J=C$ and~$I\cap J=\{0\}$
we get that $\left<q,r\right>\colon C\to C/I\,\times\,C/J$
is an isomorphism
(by the Chinese remainder theorem).\myQED
\end{myproof}

\begin{proposition}
\label{prop:cr-extensive}
The category~$\op{\CRng}$ is extensive.
\end{proposition}
\begin{myproof}
Combine Lemma~\ref{lem:cr-pushout},
Corollary~\ref{cor:cr-product-pushout-1},
and Corollary~\ref{cor:cr-product-pushout-2}.\myQED
\end{myproof}

%%%% end \auxproof of the appendox
}
%%%%

\end{document}